\documentclass[prd,11pt,onecolumn,preprintnumbers,superscriptaddress,tightenlines,nofootinbib,letterpaper,longbibliography,twoside]{revtex4-2}

\usepackage{multirow}
\usepackage{epsfig}
\usepackage{graphicx}
\usepackage{subfigure}
\usepackage{placeins}
\usepackage{mathrsfs}
\usepackage{amssymb}
\usepackage[table,xcdraw]{xcolor} 
\usepackage{colortbl} 
\usepackage{float}
\usepackage{verbatim}
\usepackage{amsmath}
\usepackage{enumitem}
\usepackage[toc,page]{appendix}
\usepackage{lmodern}
\usepackage[colorlinks=true,linkcolor=blue]{hyperref}
\hypersetup{linktocpage=true}
\usepackage[percent]{overpic}

\usepackage{multirow}

\expandafter\ifx\csname package@font\endcsname\relax\else
 \expandafter\expandafter
 \expandafter\usepackage
 \expandafter\expandafter
 \expandafter{\csname package@font\endcsname}%
\fi

\usepackage{booktabs}
\usepackage{multirow}

\begin{document}

\title{From $U(1) \times U(1)$ Symmetry Breaking to Majoron Cosmology: Insights from NANOGrav 15-year Data}
\author{Tathagata Ghosh}
\email{tathagataghosh@hri.res.in}
\affiliation{Regional Centre for Accelerator based Particle Physics, Harish-Chandra Research Institute, A CI of Homi Bhabha National Institute, Chhatnag Road, Jhunsi, Prayagraj 211019, India}
\author{Kousik Loho}
\email{kousikloho@hri.res.in}
\affiliation{Regional Centre for Accelerator based Particle Physics, Harish-Chandra Research Institute, A CI of Homi Bhabha National Institute, Chhatnag Road, Jhunsi, Prayagraj 211019, India}
\author{Sudip Manna}
\email{sudipmanna@hri.res.in}
\affiliation{Regional Centre for Accelerator based Particle Physics, Harish-Chandra Research Institute, A CI of Homi Bhabha National Institute, Chhatnag Road, Jhunsi, Prayagraj 211019, India}

\begin{abstract}
 We study the cosmology of a modified majoron model motivated by the need to protect a global $U(1)$ symmetry from gravity-induced hard explicit breaking (by $d \leq 4$ operators) at the Planck scale. The model extends the Standard Model by introducing a gauged $U(1)_{B-L}$ and an approximate global $U(1)$ symmetry, each spontaneously broken by a corresponding complex scalar singlet. This setup gives rise to a network of effectively global and local cosmic strings, whose stochastic gravitational wave signals can jointly account for the spectrum observed by the NANOGrav collaboration, particularly for majoron masses $m_{\chi} < 10^{-23}$ eV. Although the fit is not as strong as that from supermassive black hole mergers, the model still provides an alternative explanation rooted in high-energy physics. The majoron mass induces an infrared cutoff in the gravitational wave spectrum, and we examine how this effect shapes the fit to the NANOGrav data. The model also generates light neutrino masses via the seesaw mechanism and avoids cosmological constraints from $\Delta N_{\text{eff}}$, CMB anisotropies, and isocurvature fluctuations. Although the majoron can contribute to dark matter through thermal, coherent oscillation, and string-induced production mechanisms, its relic abundance remains subdominant in the NANOGrav-compatible region. In contrast, the measured dark matter relic density is achievable at higher $m_\chi$, though at the cost of tension with cosmological bounds. If the NANOGrav fits are viewed as constraints, given their comparatively lower Bayes factors, they yield bounds that are significantly stronger than those imposed by the CMB and other cosmological data.
\end{abstract}
\begin{flushright}
\small{HRI-RECAPP-2025-07}
\end{flushright}
\maketitle

\tableofcontents

\section{Introduction}
\label{sec:intro}

The first direct detection of gravitational waves (GWs) by the LIGO–Virgo collaboration in 2015~\cite{LIGOScientific:2016aoc} opened a new era in observational astronomy. In 2023, pulsar timing array (PTA) collaborations— the Parkes PTA~\cite{Reardon:2023gzh}, the European PTA~\cite{EPTA:2023fyk}, the Chinese PTA~\cite{Xu:2023wog}, and North American Nano-hertz Observatory for Gravitational Waves (NANOGrav)~\cite{NANOGrav:2023gor, NANOGrav:2023hde}—reported evidence for a stochastic GW background in the nano-Hertz range. PTAs detect such signals by monitoring the arrival-time fluctuations of radio pulses from millisecond pulsars across the galaxy~\cite{Sazhin:1978myk,1979ApJ...234.1100D,1990ApJ...361..300F}, with angular correlations following the Hellings–Downs curve~\cite{Hellings:1983fr} expected from spin-2 GWs. Among these, NANOGrav provides the most compelling evidence using 67 pulsars. While mergers of supermassive black hole binaries (SMBHBs) remain the leading astrophysical explanation~\cite{Sesana:2004sp, Burke-Spolaor:2018bvk}, no individual SMBHB mergers have been detected, motivating interest in cosmological sources~\cite{NANOGrav:2023hvm, EPTA:2023xxk}. These include domain walls~\cite{Ferreira:2022zzo,An:2023idh,Guo:2023hyp,Ellis:2023oxs}, phase transitions~\cite{Caprini:2010xv,DiBari:2021dri}, scalar-induced GWs~\cite{Vaskonen:2020lbd,Sugiyama:2020roc,Liu:2023ymk,Liu:2023pau}, and cosmic strings~\cite{Ellis:2020ena,Bian:2022tju}. In this work, we focus on cosmic strings (CSs), which not only provide a plausible source of the signal but also offer a rare probe of high-scale physics, spanning $10^{12}$–$10^{15}$ GeV (depending on local or global CS), well beyond the reach of current and proposed future colliders.

A recent work on global cosmic strings (GCSs)~\cite{Fu:2023nrn} shows that their GW spectrum, for symmetry-breaking scales around $\mathcal{O}(10^{15})$ GeV, can fall within the nano-Hertz range probed by PTAs. Intriguingly, the classic and simplest majoron model~\cite{Chikashige:1980ui}—a well-known framework for neutrino mass generation—features a global $U(1)_{L}$ symmetry broken at $\mathcal{O}(10^{14}$–$10^{15})$ GeV by a complex scalar $\phi$. This symmetry breaking has two key implications: (i) it generates GCSs that may emit GWs detectable by PTAs such as NANOGrav, and (ii) it induces high-scale Majorana masses for right-handed neutrinos (RHNs), enabling light neutrino mass generation via the Type-I seesaw mechanism~\cite{Minkowski:1977sc,Yanagida:1980xy,Gell-Mann:1979vob,Glashow:1979nm,Mohapatra:1979ia,Schechter:1980gr,Georgi:1981pg}. {The majoron is identified with the phase of the scalar field $\phi$. While the GCS-induced GW spectrum, characterised by a negative spectral tilt, provides a less favourable fit to the latest data than SMBHB mergers, it still offers a promising opportunity to explore and constrain the parameter space of the majoron model. Additionally, majoron models also have a wide range of astrophysical, cosmological, and laboratory implications~\cite{Mohapatra:1998rq}}.

{Despite its appealing features, the simplest majoron model suffers from a theoretical limitation. Gravitational effects are known to explicitly violate global symmetries at the Planck scale through wormhole-induced processes~\cite{Abbott:1989jw, Preskill:1992tc}. One can integrate out wormholes to generate an effective theory of the scalar fields charged under the global symmetry~\cite{Abbott:1989jw}. Suppose the global $U(1)$ symmetry is only softly broken by gravity and arises accidentally at the seesaw scale. In that case, the resulting majoron survives as a pseudo-Nambu–Goldstone boson (pNGB)~\cite{Rothstein:1992rh}, potentially serving as a dark matter (DM) candidate—unlike in local $U(1)_L$ or $U(1)_{B-L}$ seesaw frameworks. However, the wormhole-induced effective Lagrangian does not guarantee that a single scalar field is sufficient to forbid all dangerous $d \leq 4$ operators, which could result in hard symmetry breaking at the Planck scale~\cite{Abbott:1989jw}. Moreover, we will demonstrate that fitting the NANOGrav data within this framework leads to tension with cosmological bounds from $\Delta N_{\text{eff}}$, cosmic microwave background (CMB) anisotropies, and isocurvature perturbations.}

{To address these issues, we adopt a more robust framework, which we call the modified majoron model, by gauging the $U(1)_{B-L}$ symmetry and introducing an additional complex scalar $\phi^{\prime}$~\cite{Rothstein:1992rh}. This extended model includes both a local $U(1)_{B-L}$ and an approximate global $U(1)$ symmetry. Both $\phi$ and $\phi^{\prime}$ are charged under $U(1)_{B-L}$, with only $\phi$ carrying the global charge. Assigning $(B-L)$ charge of 2 to $\phi$ enables Majorana mass generation for RHNs. The $(B-L)$ charge of $\phi^{\prime}$ can be assigned such that the leading $U(1)_{B-L}$-invariant operator that breaks the global $U(1)$ symmetry arises only at dimension $d \geq 5$. The phase of $\phi$ still constitutes the majoron in this model and is the potential DM candidate.
While the primary focus of this work is on the cosmological implications of the modified majoron model, we also briefly compare the results with the same from the simplest majoron model.}

Recent studies have explored a broad landscape of stable CSs—including cosmic superstrings, local strings, and global strings~\cite{Schmitz:2024gds,Koehler:2024gzv,Wachter:2024zly,Wachter:2024aos,Blasi:2024vew,Hirose:2024wuk,Sousa:2024ytl,Cheng:2024axj,Baeza-Ballesteros:2024otj,Favero:2024xwh,Antusch:2024qpb,Storm:2024yhf,Chianese:2024gee,Schmitz:2024hxw,Blanco-Pillado:2024aca,Kume:2024adn,Pallis:2024mip,Jiao:2024rcr,Servant:2023tua,Baeza-Ballesteros:2023say,Fu:2023nrn,King:2023cgv,DiBari:2023mwu}. Some works focus on theoretical developments~\cite{Schmitz:2024gds,Sousa:2024ytl,Cheng:2024axj,Storm:2024yhf,Schmitz:2024hxw}, while others emphasize simulations of string networks~\cite{Baeza-Ballesteros:2024otj,Koehler:2024gzv,Wachter:2024zly,Hirose:2024wuk,Blanco-Pillado:2024aca,Jiao:2024rcr,Servant:2023tua,Baeza-Ballesteros:2023say}. Additional investigations have addressed CSs in the contexts of leptogenesis~\cite{Chianese:2024gee}, inflation~\cite{Antusch:2024qpb,Pallis:2024mip}, and dark matter phenomenology~\cite{Favero:2024xwh}. {Ref.~\cite{Fu:2023nrn} compared the GW spectrum from a global $U(1)_{B-L}$ symmetry breaking against the NANOGrav 12.5 year data. The GW spectrum from GCSs in a majoron model is also considered in Ref.~\cite{DiBari:2023mwu}. However, neither paper considers the impact of gravity-induced soft breaking of the global symmetry. On the other hand, Ref.~\cite{Kume:2024adn} presents a Bayesian analysis of the NANOGrav 15-year data to place an upper bound on the string tension of local cosmic strings (LCSs).
}

The extended symmetry structure of the modified majoron model gives rise to the formation of intricate topological defects in the early Universe. When the local $U(1)_{B-L}$ symmetry breaks, it generates LCSs, followed by the formation of GCSs from the spontaneous breaking of the global $U(1)$. As the Universe cools down to a temperature where the Hubble constant is comparable to the majoron mass, explicit symmetry breaking becomes relevant, leading to the formation of domain walls (DWs). Each of these topological defects can emit gravitational radiation. The string-wall network's fate hinges on how many walls attach to each string. If stable DWs form, they can dominate the Universe's energy density, contradicting the CMB observations. However, if only a single DW attaches to either type of string, the network becomes unstable and promptly collapses~\cite{Rothstein:1992rh, Barr:1992qq, Vilenkin:1982ks}. In this paper, we analyse scenarios where only a single DW is formed so that the GW spectrum is generated predominantly from CSs.\footnote{In particular, we consider the scenario in which one DW is attached to one GCS. In this case, minimization of Majoron gradient energy around the local strings suggests that no wall can consistently attach to a local string (see section~\ref{gcs+lcs} for details), rendering the system unstable and leading to the eventual collapse of the domain wall~\cite{Vilenkin:1982ks}. By contrast, when a single DW is attached to a local string, more than one wall can be attached to the associated global string. To the best of our knowledge, the collapse of this particular complex wall-string network has not been explicitly demonstrated in dedicated numerical simulations. Given this absence of direct numerical confirmation, we restrict our analysis to the case of a single domain wall attached to global strings only; see Table~\ref{tab:BP} for details.}
An interesting aspect explored in this work is the role of the majoron mass in setting an infrared (IR) cutoff on the GW spectrum, which in turn influences the fit to the NANOGrav data.

Recently, Ref.~\cite{Greljo:2025suh} constructed a model gauging an anomaly-free linear combination of baryon and lepton numbers to ensure proton stability. This model also features a gravity-induced, explicitly broken accidental $U(1)_L$ global symmetry and the associated majoron. The authors explore the majoron as a DM candidate for $m_{\chi} > 10^{-5}$ eV and consider minimal thermal leptogenesis. Our fit to the NANOGrav data extends to this class of model as well. However, we focus on majoron masses below $10^{-15}$ eV, necessary for the string-wall network to survive long enough for the CS GW spectrum to reach below the nano-Hz frequency range.

The paper is organised as follows: Section~\ref{sec:model} outlines the Simplest majoron model and the modified model inspired by the short-distance global symmetry breaking generated due to gravitational effects. In Section~\ref{sec:CS}, we discuss the generation of GW spectra from GCSs, LCSs, and strings arising in the modified model. Section~\ref{sec:results} presents our fits to the NANOGrav 15-year data using the GW spectra from both the simplest and the modified majoron model. Cosmological and astrophysical constraints on the NANOGrav-preferred parameter spaces are examined in Section~\ref{sec:cons}. The viability of the majoron as a DM candidate is explored in Section~\ref{sec:majoronDM}, followed by the summary and concluding remarks in Section~\ref{sec:conc}.

\section{Model}
\label{sec:model}
\subsection{The simplest majoron model}
\label{sec:themajoronmodel}

The minimal majoron model consists of an additional $U(1)_L$ (lepton number) global symmetry on top of the SM gauge groups~\cite{Chikashige:1980ui}.
The particle content of the model is extended by three singlet RHNs, $N_I$, and a complex scalar singlet  $\phi$.
The spontaneous breaking of this global $U(1)_L$ symmetry generates Majorana mass terms for the RHN fields. 
For the sake of simplicity, we assume that there is no tree-level interaction term between the SM Higgs doublet, $\Phi$, and the singlet scalar $\phi$. 
Hence, the additional tree-level contribution to the Lagrangian in the simplest majoron model, beyond the Standard Model component, can be expressed as
\begin{equation}
\mathcal{L}_{\text{new}} = \frac{i}{2}\overline{N_I}\gamma^\mu\partial_\mu N_I+\frac{1}{4}\lambda_{IJ}\overline{N_I}(1+\gamma_5)N_J\phi+\frac{1}{4}\tilde{\lambda}_{IJ}\overline{\nu_I}(1+\gamma_5)N_J\tilde{\Phi}-V_1(\phi)+ \text{h.c.}
\label{eq.1}
\end{equation}
where $\tilde{\Phi}$ is the dual of the SM Higgs doublet, and $N_I, \nu_J$ are 4-component Majorana spinors (i.e $N^c_I = N_I$ and $\nu^c_J = \nu_J$).
The indices $I, J = 1, 2, 3$ are the generation indices. $V_1(\phi)$ is the tree-level potential of the singlet scalar $\phi$. In general, the Majorana-type Yukawa coupling matrix $\lambda$ can contain off-diagonal entries. However, for simplicity, we again assume the $\lambda$ matrix to be diagonal (hereafter, $\lambda_{II}$ will be simplified to $\lambda_I$). In this model, the lepton number is a conserved charge since all the SM lepton doublets and RHNs possess $U(1)_L$ charge of $L = 1$ and  $\phi$ carries $L = -2$. Consequently, $\phi$ interacts with RHNs only.

Prior to the breaking of the $U(1)_L$ global symmetry, all RHNs were massless. Once $\phi$ acquires a vacuum expectation value (VEV), $\eta$, to spontaneously break the global symmetry, it generates high-scale Majorana masses to RHNs. Assuming $V_1(\phi)=\lambda\bigg(\phi^\star\phi-\frac{\eta^2}{2}\bigg)^2 $, after spontaneous symmetry breaking, the field $\phi$ can be expressed as
\begin{align}
    \phi = \frac{(\rho+\eta)}{\sqrt{2}}\exp{(i\chi/\eta)}.
    \label{eq.2}
\end{align}
Here, the $CP$-odd component, $\chi$, is a massless Nambu-Goldstone boson called majoron, and the $CP$-even component, $\rho$, is a massive scalar with mass $m_\rho = \sqrt{2\lambda}\eta$. Additionally, the nonzero VEV of $\phi$ generates Majorana mass terms for RHNs, $M_I=\lambda_{I}\eta/\sqrt{2}$. 
One gets rid of the majoron field from the neutrino mass matrix by rotating the neutrino fields in the following manner:
$N_I\rightarrow\exp{(-i\chi\gamma_5/2\eta)}N_I$ and $\nu_I\rightarrow\exp{(i\chi\gamma_5/2\eta)}\nu_I$. Furthermore, these rotations induce the following interaction terms between the majoron and the neutrinos of the theory: 
$\mathcal{L}_\text{new}\supset\frac{(\partial_\mu\chi)}{4\eta}(\overline{\nu_I}\gamma^\mu\gamma_5\nu_I-\overline{N_I}\gamma^\mu\gamma_5N_I)$. The spontaneous breaking of the electroweak symmetry as the SM Higgs doublet develops a VEV ($v_{ew}$) leads to the generation of a Dirac mass term: $(m_D)_{IJ}=\tilde{\lambda}_{IJ}v_{ew}/\sqrt{2}$, which takes part in the Type-I seesaw mechanism~\cite{Minkowski:1977sc,Yanagida:1980xy,Gell-Mann:1979vob,Glashow:1979nm,Mohapatra:1979ia,Schechter:1980gr,Georgi:1981pg} along with the lepton number violating RHN Majorana mass terms to generate the tiny masses for the light neutrinos given as
\begin{equation}
(m_\nu)_{IJ}\sim\frac{\tilde{\lambda}_{IK}\tilde{\lambda}_{KJ}v_{ew}^2}{2\sqrt{2}\lambda_K\eta} = \frac{\delta_{IJ}v_{ew}^2}{2\sqrt{2}\eta} \, ,
\label{eq.3}
\end{equation}
%
where $v_{ew} $ = 246 GeV. The spontaneous symmetry breaking will also lead to the formation of GCSs in the early Universe. These GCSs will radiate gravitationally and produce detectable signals at PTA experiments, as we shall discuss in Section~\ref{gcs}.

The minimal majoron model not only solves one of the unsolved issues of the SM and generates tiny neutrino masses but also provides a dynamical mechanism to generate large Majorana masses for the RHNs. However, it is a well-known argument that gravitational effects are expected to break global symmetries explicitly at the Planck scale~\cite{Abbott:1989jw, Preskill:1992tc}. An obvious fix is to elevate the global $U(1)_L$ symmetry to a local one. However, we opt for a different approach in which the global $U(1)$ symmetry appears as an accidental one at the see-saw scale and is broken softly by interaction with gravity. The additional benefit of this model is that the majoron remains present in the theory as a pNGB and can be a potential DM candidate, which is another outstanding issue within the SM. In the next subsection, we discuss the construction of this modified majoron model.

\subsection{The modified majoron model}
\label{sec:themodifiedmajoronmodel}

As discussed above, we adopt a model for our analysis in which the majoron behaves as a pNGB and remains in the low-energy theory due to an accidental global $U(1)$ symmetry. Gravitational interaction introduces a soft breaking of this symmetry through higher-dimensional operators ($d > 4$), which arise due to the formation of wormholes.
In the literature, it is a common practice to express these higher-dimensional operators using only a single scalar field~\cite{Reig:2019sok}. However, following the wormhole solution presented in Ref.~\cite{Abbott:1989jw}, integrating out wormholes leads to an effective low-energy theory given by
\begin{equation}
    \mathcal{L}_{\text{eff}} = \sum_{d=1}^{\infty} \alpha_d g_d \, \phi^d  +  \text{h.c.} \,,  
    \label{eq:L_wh_eff}
\end{equation}
where $\alpha_d$ are complex parameters that correspond to the vacuum state of the theory and
\begin{equation}
    g_d \approx \left( \frac{\lambda^{(d-4)/3} d^{(d-16)/3}}{M_{Pl}^{(d-4)}} \right) e^{-(\lambda^{1/3} d^{4/3})}= \frac{\tilde{g}_{d,\text{grav}}}{M_{Pl}^{(d-4)}} \, .
\end{equation}
The reduced coupling $\tilde{g}_{d,\text{grav}}$ is exponentially suppressed by the combined scalar and gravitational Euclidean action in the region near the wormhole and is expected to be very small. For $d = 5-20$ and  $\lambda=1$, $\tilde{g}_{d,\text{grav}} \sim 10^{-(d+2)}$.
Given the structure of the effective Lagrangian in Eq.~(\ref{eq:L_wh_eff}), it is unclear whether a single scalar field can suppress $d \leq 4$ effective operators and thus prevent explicit hard breaking of the global symmetry at the Planck scale.

To avoid this lack of clarity, we prefer a conservative approach and choose a model where a local gauge symmetry ensures that no $d \leq 4$ effective operators are induced by gravity. Such a model was constructed in Ref.~\cite{Rothstein:1992rh} by gauging $U(1)_{B-L}$ and introducing an additional complex scalar field, $\phi^{\prime}$. The charge assignment of $\phi^{\prime}$ under $(B-L)$ can be chosen such that the lowest-dimensional $U(1)_{B-L}$-invariant operator that breaks the global $U(1)$ symmetry appears only at $d \geq 5$.

In a nutshell, this model extends the SM gauge group by introducing a local $U(1)_{B-L}$ symmetry and an approximate global $U(1)$ symmetry. The particle content is extended by three RHNs and two complex scalars, $\phi$ and $\phi^{\prime}$, which are all singlets under the SM gauge groups. Here, both $\phi$ and $\phi^{\prime}$ are charged under the local $U(1)_{B-L}$ symmetry, and only $\phi$ is charged under the accidental global symmetry. If one assigns $(B-L)$ charge of $q =2$ to $\phi$, it again couples with RHNs and generates Majorana masses for them. On the other hand, we keep $q^{\prime}$, the $(B-L)$ charge of $\phi^{\prime}$, as general as possible (including fractional charges) to allow us to construct different models for a given $d$-dimensional soft global symmetry breaking operator.

The lowest dimensional quantum gravity induced global symmetry breaking, but the $(B-L)$ invariant operator is 
\begin{align}
   \mathcal{L_{\text{grav}}} = g_{\text{grav}} \frac{{\phi}^{n}{(\phi^{\prime*})}^{n^\prime} }{M_{Pl}^{d-4}} + \text{h.c.}
    \,,
    \label{eq.4}
\end{align}
where, $n + n^\prime = d$, and gauge invariance imposes a condition $qn -q^\prime n^\prime=0$ with $\{n, n^\prime\}
\in \mathbb{Z}^+$. To ensure that the lowest dimensional operator has $d>4$ we require $aq-bq^{\prime} \neq 0$, if $a+b=d \leq 4$ and $\{a, b \} \in \mathbb{Z}^+$. Hence, the scalar potential of the model involving the singlet scalar fields is given by 
\begin{align}
V_1(\phi, \phi^\prime)=-\tilde{\mu}^2 |\phi|^2 + \lambda|\phi|^4 -\tilde{\mu}^{\prime2} |\phi^\prime|^2 + \lambda^\prime|\phi^\prime|^4 + \lambda_{\text{int}}|\phi|^2|\phi^\prime|^2 + \Bigg( g_{\text{grav}} \frac{{\phi}^{n}{(\phi^{\prime*})}^{n^\prime} }{M_{Pl}^{d-4}} + \text{h.c.} \Bigg)
    \,,
    \label{v_phi_phi_p}
\end{align}
where all the coefficients are positive, and $\phi = \frac{(\rho+\eta)}{\sqrt{2}}\exp{(iJ/\eta)}$ and $\phi^{\prime} = \frac{(\rho^{\prime}+\eta^{\prime})}{\sqrt{2}}\exp{(iJ^{\prime}/\eta^{\prime})}$.
Once $\phi^{\prime}$ acquires a VEV ($\eta^{\prime}$), it spontaneously breaks the $U(1)_{B-L}$. If $\eta^{\prime} > \eta$, the global $U(1)$ symmetry is still preserved up to a tiny soft breaking arising from the operator of Eq.~(\ref{eq.4}). Eventually, when the field $\phi$ develops a VEV ($\eta$), the global $U(1)$ is also spontaneously broken, giving rise to Majorana masses for RHNs similar to Section~\ref{sec:themajoronmodel}. 

Once the $U(1)_{B-L}$ charge of $\phi$ is fixed at $q=2$ from the requirement of Majorana mass generation of RHNs, one is left with the freedom of only one parameter choice at their disposal to introduce a global symmetry breaking term in dimension $d$. One can thus choose to vary the parameter $n$, which can admit positive integer values up to $d-1$, and the values of $n^\prime$ and $q^\prime$ can be updated accordingly using the dimensional and gauge invariance conditions mentioned above. Such terms in the Lagrangian will inevitably give rise to an $M_{Pl}^{d-4}$-suppressed majoron mass~\cite{Rothstein:1992rh}. 
For example, in dimension-5 for $q=2$, the $(B-L)$ invariant but global symmetry breaking term can arise if $q^\prime$ takes values \{1/2, 4/3, 3, 8\}. In contrast, if we assign a fractional $U(1)_{B-L}$ charge to $\phi^{\prime}$, say $q^\prime=1/8$, then the lowest dimension at which the global symmetry can break explicitly is 17. More phenomenologically motivated benchmark points are shown in Table~\ref{tab:BP_DM} in Section~\ref{sec:majoronDM}.

After symmetry breaking, a linear combination of {$CP$-odd scalars} 
is eaten by the $Z^{\prime}_{B-L}$ gauge boson, and the orthogonal component given by
\begin{align}
   \chi = \frac{q^{\prime} \eta^{\prime} J - q \eta J^{\prime}}{\sqrt{q^2 \eta^2 + {q^{\prime}}^2 {\eta^{\prime}}^2}} 
    \label{eq.majoron}
\end{align}
becomes the majoron in this theory. The explicit symmetry-breaking term will induce a majoron mass
\begin{align}
   m_{\chi}^2 = \frac{ g_{\text{grav}}}{2^{(\frac{d}{2}-1)}} \frac{\eta^n {\eta^{\prime}}^{n^\prime}}{M_{Pl}^{d-4}} \frac{n n^{\prime}}{q q^{\prime} \eta_{\text{eff}}^2}
    \,,
    \label{eq.majoronMass}
\end{align}
where 
\begin{equation}
\eta_{\text{eff}}^2= \frac{\eta^2 {\eta^{\prime}}^2}{{q^2 \eta^2 + {q^{\prime}}^2 {\eta^{\prime}}^2}}.
\label{eq:etaeff}
\end{equation}

This framework will enable majoron to be a pNGB candidate, which can play an important role in early Universe cosmology (nucleosynthesis, coherent oscillation of pNGB around its minima, etc.). Most of the phenomenology of this majoron will be governed by its interaction with light neutrinos, which is given by, 
\begin{align}
   \mathcal{L}_{\nu \nu \chi} = \frac{q^{\prime} \eta_{\text{eff}}}{4 \eta^2} (\partial_{\mu} \chi) (\overline{\nu}_i \gamma^{\mu} \gamma_5 \nu_i) \, \, (i=1,2,3).
    \label{eq.ChiNuInteraction}
\end{align}
majoron can couple to other fermions via $Z$ boson and light neutrino-mediated loops. For further details of the model and exhaustive phenomenological discussions, we refer the readers to Ref.~\cite{Rothstein:1992rh}.

In the early Universe, majorons can be produced thermally and also non-thermally from coherent oscillations and topological defects~\cite{Rothstein:1992rh, Reig:2019sok, Gorghetto:2021fsn}. Hence, it can become a DM candidate if its lifetime is longer than the age of the Universe. We discuss the DM aspects of majoron in this model in Section~\ref{sec:majoronDM}. Additionally, the complex symmetry structure of the model gives rise to the formation of intricate topological defects in the early Universe. First, when the temperature of the universe $T \sim \eta^{\prime}$, $U(1)_{B-L}$ symmetry breaking generates a set of LCSs. Next, once the global $U(1)$ symmetry is broken at $T \sim \eta$ global strings form. When the Universe cools further to a temperature $T_{*}$ such that $m_{\chi} = 3 H(T_{*})$, then the effect of explicit symmetry breaking becomes dynamically important, and DWs form. All these topological defects can radiate gravitationally, which we discuss in Section~\ref{gcs+lcs}.

\section{Cosmic Strings and Gravitational Waves}
\label{sec:CS}

Cosmic strings~\cite{VILENKIN1985263, Hindmarsh:1994re, Vilenkin:2000jqa} are one of the topological defects that can arise in the early Universe when a symmetry is spontaneously broken by a complex scalar field. For the formation of CSs, the scalar field must have a vacuum manifold $\mathcal{M}$ so that the fundamental (or the first homotopy) group of $\mathcal{M}$ is non-trivial, i.e., $ \pi_1 (\mathcal{M}) \neq I$.
A theory containing a complex scalar field $\phi$ with a quartic `Mexican hat' potential, characterized by a degenerate circle of minima $|\phi| = \eta/\sqrt{2}$, satisfies the necessary condition for cosmic string formation.
These CSs can be present as remnants of the symmetry breaking in the broken phase of the Universe.

If there is no energy-loss mechanism for the strings, their energy density scales as $1/a^2$, and they would have gradually dominated the total energy budget of the Universe. However, that is inconsistent with the CMB data obtained by Planck~\cite{Planck:2013pxb}.
{Nevertheless, if there is an efficient energy loss mechanism for the CSs and if that can eventually lead the CS's characteristic length (correlation length) scale $L$ to become proportional to the cosmological (Hubble) scale, which is also known as reaching the scaling regime,} 
then such a scenario can be compatible with the observed CMB data. 
The only plausible mechanism~\cite{Turok:1984db} for reaching the scaling regime is the production of closed loops through the collision of two strings and/or self-intersection of strings and their subsequent decay via oscillation in their fundamental mode and higher harmonics and radiating gravitational waves (GWs) and/or Goldstone-bosons. Over time, CSs can potentially form a network which consists of horizon-sized long strings and sub-horizon-sized loops.
In addition to energy emission from the normal modes of loop oscillation, there can be local deformations called cusps and kinks, which can produce bursts of GWs. Cusps are the points in a string loop that occur once per oscillation period, where the string motion 
reaches the speed of light, and kinks are the discontinuities in the tangent vector of a string-loop.  

The standard form of the GW relic energy density is given by 
\begin{align}
 \Omega_{\text{GW}}=\frac{f}{\rho_c}\frac{d\rho_{\text{GW}}}{df}
    \,,
    \label{eq.17}
\end{align}
where $\rho_{GW}$ is the GW energy density and $\rho_c=(\sqrt3H_0)^2/8\pi G$ is the critical energy density. We can decompose the total GW radiation from a loop into a sum of normal-mode oscillations with oscillation frequency $ \tilde{f}= 2k/\tilde{l}$, where $k \in \mathbb{Z}^+$. Here, $\tilde{l}\equiv \tilde{l}(\tilde{t})$ is the instantaneous loop-size at original GW-emission time, $\tilde{t}$. Taking the redshift into account, the observed frequencies today can be written as 
\begin{align}
f_k=\tilde{f}_k\frac{a(\tilde{t})}{a(t_0)}=\frac{2k}{\tilde{l}}\frac{a(\tilde{t})}{a(t_0)}
    \,,
    \label{eq.18}
\end{align}
where {$a$ is the scale factor}, and $t_0$ represent the present time. Summing over all harmonic modes, the total GW relic density will be of the form 
\begin{align}
\Omega_{\text{GW}}(f)= \sum_k \Omega_{\text{GW}}^{(k)}(f)=\sum_k\frac{f_k}{\rho_c}\frac{d}{df_k}(\rho_{\text{GW}})
    \,.
    \label{eq.19}
\end{align}

Depending on the global or local nature of the underlying symmetry group of the theory, one can classify the CSs into global strings and local strings accordingly. In the following subsections, we estimate the GW spectrum using Eq.~(\ref{eq.19}), starting with global strings, then local strings, and finally strings formed in the modified majoron model.

\subsection{Gravitational Waves from global cosmic strings}
\label{gcs}

Global cosmic strings can form when a $U(1)$ global symmetry is spontaneously broken. For GCS, most of the energy loss happens due to Goldstone boson radiation, and GW emission is only a secondary source of radiation. The amplitude of the GCS-mediated GW spectrum is usually much suppressed compared to the Nambu-Goto LCS-induced GW spectrum.
In this subsection, we discuss GCS-induced GW-spectrum~\cite{Chang:2021afa, DiBari:2023mwu}.

When CSs reach the scaling regime, the inter-string separation scale (correlation length) $L$ remains nearly constant, and the energy density of the GCS network (primarily stored within long strings) can be written as
\begin{align}
    \rho_{\text{GCS}} = \frac{\mu(t)}{L^2(t)} = \mu(t) \frac{\xi(t)}{t^2}
    \,,
    \label{eq.5}
\end{align}
where $\xi(t)$ stands for the number of long strings per horizon volume. $\mu(t)$ in the above equation denotes the long strings' energy per unit length given by, 
\begin{align}
    \mu(t) =\pi\eta^2+2\pi \eta^2 \log \frac{L(t)}{\delta}\simeq 2\pi \eta^2 \log \frac{L(t)}{\delta} \equiv 2 \pi \eta^2 N(t)
    \,,
    \label{eq.6}
\end{align}
where $\eta$ is the symmetry breaking scale, $\delta(t)\sim \frac{1}{\lambda \eta}$ corresponds to the width of the GCS core, with $\lambda $ being the quartic coupling of the scalar potential and  N(t) $\equiv$  $\log [L(t)/\delta] $ is a time-dependent dimensionless parameter. The logarithmic divergence of $\mu(t)$ is the dominant term in Eq.~(\ref{eq.6}), which appears due to a long-range force induced by the Goldstone boson of the global theory. {These strings intersect each other and/or self-intersect and chop off loops, which eventually form a string network and radiate Goldstone bosons and GWs.}

Assuming that the initial size of the loops at the time of their formation ($t_i$), $l_i=\alpha t_i$ is nearly monochromatic ($\alpha \sim 0.1 $), where $\alpha$ is a dimensionless scaling parameter, the rate of change of the energy density of the loops in a scaling network can be expressed as
\begin{align}
  \frac{d\rho_0}{dt}=-\frac{d\rho_{\textrm{GCS}}}{dt}\times{F}_{\alpha}=\varepsilon_{\text{loop}} \frac{\mu}{t^3}{F}_{\alpha}
    \,,
    \label{eq.13}
\end{align}
where $\rho_0$ stands for the energy density of string loops. On the other hand, $F_\alpha \sim 0.1$ denotes the proportion of energy contained in loops that can be emitted as radiation, such as gravitational waves or Goldstone bosons.
$\varepsilon_{\text{loop}}\equiv \langle c \rangle v(t) \xi^{3/2}$ is the loop emission parameter, where $ v(t)$ is the average string velocity, $\langle c \rangle$ is is the loop chopping rate and  $\xi(t)=\frac{t^2}{L^2(t)}$. The form of $L(t)$ and $v(t)$ can be estimated using the velocity-dependent one scale (VOS) model~\cite{Chang:2021afa}. With the help of the simulation results~\cite{Gorghetto:2018myk, Hindmarsh:2019csc, Klaer:2017qhr}, we can approximate the value of $\langle c \rangle=0.497$. 

After formation, GCS loops oscillate and radiate energy via Goldstone boson and GW emission until they disappear completely. The rate of energy loss of the loops can be approximated as~\cite{Vilenkin:1986ku},
\begin{align}
  \frac{dE}{dt}=\Gamma G\mu^2-\Gamma_g \eta^2
    \,,
    \label{eq.14}
\end{align}
where the first and second term on the right-hand side denotes the loss in loop energy due to GW and Goldstone boson emission, respectively. Dedicated simulations have been carried out for the best-approximated values of $\Gamma$ and $\Gamma_g$ and the suggested benchmark values are $\Gamma \sim 50$~\cite{Blanco-Pillado:2013qja, Blanco-Pillado:2017oxo, Vilenkin:1981bx, Blanco-Pillado:2011egf} and $\Gamma_g \sim 65$~\cite{Battye:1997jk, Vilenkin:2000jqa}. Therefore, the loop size with initial length $l_i=\alpha t_i$ in light of Eq.~(\ref{eq.14}) can be expressed at a later time as
\begin{align}
  l(t)\simeq \alpha t_i -\Gamma G\mu(t-t_i)-\frac{\Gamma_g}{2\pi}\frac{t-t_i}{ N}
    \,,
    \label{eq.15}
\end{align}
where $t_i$ stands for the time of the loop formation. Considering radiation from cusps as the dominating channel of GW and Goldstone emission~\cite{Olum:1998ag, Blanco-Pillado:2015ana, Blanco-Pillado:2019nto}, the radiation parameters in the Fourier basis can be written as
\begin{align}
\Gamma^{(k)} &= \frac{\Gamma k^{-\frac{4}{3}}}{\sum_{n=1}^{\infty} n^{-\frac{4}{3}}}, \quad \text{and} \quad
\Gamma_g^{(k)} = \frac{\Gamma_g k^{-\frac{4}{3}}}{\sum_{n=1}^{\infty} n^{-\frac{4}{3}}} \,,
\label{eq.16}
\end{align}
where $\sum_k \Gamma^{(k)}=\Gamma$, and $\sum_k \Gamma_g^{(k)}=\Gamma_g$. The radiation from other channels like kinks, kink-kink collisions, and fundamental modes of vibrations have different power laws~\cite{Vachaspati:1984gt, Burden:1985md, Garfinkle:1987yw}.

Using Eq.~(\ref{eq.13}) and Eq.~(\ref{eq.15}) and integrating over $\tilde{t}$, we can get the  GW contribution from individual $k$ mode for GCS as~\cite{Chang:2021afa} 
\begin{align}
\Omega_{\text{GW}}^{(k)}(f)\Bigg|_\text{GCS} = \frac{\mathcal{D}_\alpha}{\rho_c}\frac{2k}{f}\frac{F_{\alpha}}{\alpha}\int_{t_F}
^{t_0}d\tilde{t}\frac{\Gamma^{(k)}G\mu^2}{ \left(\alpha+\Gamma G\mu+\frac{\Gamma_g}{2\pi N}\right)} \frac{\varepsilon_\text{loop}\left(t_i^{(k)}\right)}{{t_i^{(k)}}^4}\theta(\tilde{l}) \theta(\tilde{t}-t_i) {\left(\frac{a(\tilde{t})}{a(t_0)}\right)}^5{\left(\frac{a(t_i^{(k)})}{a(\tilde{t})}\right)}^3
\,,
\label{eq.20}
\end{align}
where $t_F$ is the string-network formation time, and the loop size distribution function is given by $\mathcal{D}_{\alpha} \sim 1$. The Heaviside theta functions ensure energy conservation and causality. A loop responsible for emitting gravitational waves at time $\tilde{t}$, corresponding to an observed frequency $f$, originated at an earlier formation time $ t_i^{(k)}(\tilde{t},f)=\left[\tilde{l}(\tilde{t},f,k)+\Gamma G \mu \tilde{t}+(\Gamma_g/2\pi N)\tilde{t} \right]/\left[ \alpha+\Gamma G \mu+ \Gamma_g/2\pi N\right] $.
The total GW relic is evaluated by numerically integrating Eq.~(\ref{eq.20}) and summing to very large values of $k$ so that the series in Eq.~(\ref{eq.19}) converges.

The GCS-mediated GW-spectrum has a vast frequency range, from $f_0 \simeq2/ \alpha t_0 \sim 3.6 \times 10^{-16} $ Hz to a very large frequency depending on the symmetry breaking scale. For our purpose, we are interested in $f_0 <f <f_{th}$
(with $f_{th}\sim 1.8 \times10^{-7} $ Hz) as PTA experiments like NANOGrav are sensitive to this range. The spectrum falls as $f^{-1/3}$ in this range and can be approximated by~\cite{Chang:2021afa} 
\begin{align}
h^2\Omega_{\text{GW}_\text{GCS}}(f) \simeq 2.9 \times10^{-12}{\left(\frac{\eta}{10^{15}\text{GeV}}\right)}^4 {\left(\frac{f}{f_{th}}\right)}^{-1/3}
    \,.
    \label{eq.21}
\end{align}
We will use only this Eq.~(\ref{eq.21}) in our subsequent GCS analysis of this paper.

\subsection{Gravitational Waves from local cosmic strings}
\label{lcs}
Local cosmic strings can form due to the breaking of a local $U(1)$ gauge symmetry. These LCSs also can produce CS networks and radiate GWs. The key difference between GCSs and LCSs is that LCSs do not have goldstone boson radiation, and they radiate energy predominantly via GW emission. 

Thus, the mass per unit length of LCS, $\mu$, has no logarithmic divergence part and becomes time-independent. In the same spirit of GCS, $\mu$ and the energy density of LCS can be written as
\begin{align}
    \mu(t) =\mu=\pi \eta^2, \quad \text{and} \quad
    \rho_\text{LCS} = \frac{\mu}{L^2(t)} = \mu \frac{\xi(t)}{t^2}
 \,.
 \label{eq.22}
\end{align}
Since the energy loss of LCS loops takes place mostly via GW emission, and no Goldstone emission mode is present, we can reformulate Eq.~(\ref{eq.14}) and Eq.~(\ref{eq.15}) for LCS in the following manner,
\begin{align}
\frac{dE}{dt}=\Gamma G \mu^2,\quad \text{and} \quad
l(t)\simeq \alpha t_i -\Gamma G\mu(t-t_i)
    \,.
    \label{eq.23}
\end{align}
Again, the total GW emission can be decomposed into a sum of normal mode oscillation. Following Eq.~(\ref{eq.20}), We can express the relic density of GWs from the $k$-th mode of LCS loops as 
\begin{align}
\Omega_{\text{GW}}^{(k)}(f)\Bigg|_\text{LCS}= \frac{\mathcal{D}_\alpha}{\rho_c}\frac{2k}{f}\frac{F_\alpha}{\alpha}\int_{t_F}
^{t_0}d\tilde{t}\frac{\Gamma^{(k)}G\mu^2}{ \left(\alpha+\Gamma G\mu\right)} \frac{\varepsilon_\text{loop}\left(t_i^{(k)}\right)}{{t_i^{(k)}}^4}\theta(\tilde{l}) \theta(\tilde{t}-t_i) {\left(\frac{a(\tilde{t})}{a(t_0)}\right)}^5{\left(\frac{a(t_i^{(k)})}{a(\tilde{t})}\right)}^3
\,,
\label{eq.24}
\end{align}
where $t_i^{(k)}(\tilde{t},f)$ for LCS is defined as
\begin{align}
 t_i^{(k)}(\tilde{t},f)=\frac{(\tilde{l}(\tilde{t},f,k)+\Gamma G \mu \tilde{t})}{( \alpha+\Gamma G \mu)}
    \,.
    \label{eq.25}
\end{align}
Here, $\Gamma^{(k)} $ is proportional to $ k^{-\tilde q}$ and $\tilde q=\{4/3, 5/3, 2\}$ corresponding to GW emission from cusps, kinks and kink-kink collisions, respectively. For more details about benchmark values of the parameters $\langle c \rangle$, $q_v$, $\alpha$, $F_{\alpha}$ and $\mathcal{D_{\alpha}}$ for LCSs we refer the reader to Ref.~\cite{Sousa:2024ytl}.

For completeness, We provided a rough sketch of how one can numerically derive the GW spectrum for LCS in Eq.~(\ref{eq.24}). However, in practice, for our numerical analysis, we have utilised several simulation templates provided in the \texttt{PTArcade} package~\cite{Mitridate:2023oar}. Of course, these templates can be generated after doing the $d\tilde{t}$ integration of Eq.~(\ref{eq.24}). Within \texttt{PTArcade}, one can find four different GW templates for LCS, namely,  \texttt{stable-k}, \texttt{stable-m}, \texttt{stable-n} and \texttt{stable-c}. The \texttt{stable-k}~\cite{Damour:2001bk} (\texttt{stable-c}~\cite{PhysRevD.31.3052}) template assumes that kinks (cusps) on closed loops serve as the source of GW emission, whereas \texttt{stable-n}~\cite{Blanco-Pillado:2011egf, Blanco-Pillado:2015ana}
considers a combined GW spectrum arising from both cusps and kinks. In contrast, the \texttt{stable-m} template focuses on GW emission from the fundamental vibrational mode of closed loops~\cite{NANOGrav:2023hvm}.

We should mention here that in \texttt{PTArcade}, the quantity $G\mu$ is used as an input parameter for GW spectra of the four templates mentioned above. A logarithmic uniform prior is used for $G\mu$. However, we should reiterate that in this paper, we are trying to constrain the symmetry-breaking scale $\eta^\prime$. Hence, we use a logarithmic uniform prior for $\eta^\prime$ and convert $\eta^\prime$ to $G\mu$ using the first relation of Eq. (\ref{eq.22}) to generate LCS GW spectra. 

The GW spectrum for LCS (GCS) presented in Eq.~(\ref{eq.24}) \big(Eq.~(\ref{eq.20})\big) is derived under the assumption that a single gauged (global) $U(1)$ symmetry spontaneously broken by an SM singlet complex scalar. However, in our modified majoron model, not only the SM gauge group is extended by a local $U(1)_{B-L}$ and a global $U(1)$ symmetry but also by the inclusion of two SM singlet complex scalar fields. Consequently, the GW spectrum from this model is expected to be more intricate and intriguing, as discussed in the following subsection.

\subsection{Gravitational Waves from cosmic strings in the modified majoron model}
\label{gcs+lcs}

As mentioned above, the formation of topological defects in the modified majoron model can be pretty involved. 
Thus, before providing the GW spectrum for this model, we take a detour and provide a summary of the complexities of topological defect formation in this model.

Under our setup, the cosmic history of the model is as follows. When $T \sim \eta'$, $\phi^\prime$ acquires a VEV ($\eta^\prime$) to break the local $U(1)_{B-L}$ symmetry. Hence, LCSs are formed, which we call Type-$\phi^{\prime}$ strings. Then, as the Universe cools down to $T \sim \eta$, the global $U(1)$ symmetry breaks, leading to the emergence of GCSs and Goldstone bosons. We refer to this network of GCSs as Type-$\phi$ strings. At the same time, where Type-$\phi^{\prime}$ LCSs are formed, the phase of $\phi$ adjusts itself by changing its winding number to minimise the kinetic energy of LCSs. As the Universe continues to cool to $T_{*}$ so that $m_{\chi} = 3 H(T_{*})$, the impact of explicit symmetry breaking becomes dynamically significant, leading to the formation of DWs~\cite{Gorghetto:2021fsn}~\footnote{Provided $m_{\chi} < H(T=\eta)$~\cite{Rothstein:1992rh}, with $H$ being the Hubble constant. Otherwise, DWs will appear at $T = \eta$ itself.}. In this paper, we focus on the $\eta<\eta^\prime$ scenario only. 
Also, it is instructive to mention here that if $n$ and $n^{\prime}$ are relatively prime, then the gravity-induced operator in Eq.~(\ref{eq.4}) breaks the global $U(1)$ symmetry to a unique vacuum~\cite{Barr:1992qq}. 

The evolution of the string-wall network is determined by the number of walls attached to different types of strings. If for a particular choice of $n$ and $n^{\prime}$, a stable DW is formed, then it will eventually dominate the energy density of the Universe, and the model has a `domain wall problem'. However, if only one DW is attached to either the Type-$\phi$ or Type-$\phi^{\prime}$ CSs, then the string-wall network collapses~\cite{Rothstein:1992rh, Barr:1992qq, Vilenkin:1982ks}. Hence, to ensure that the GW signal in the PTA frequency range originates from CSs, one needs to find the condition that either the Type-$\phi$ or Type-$\phi^{\prime}$ strings form with only one DW attached to it.

The number of maxima in the scalar potential of Eq.~(\ref{v_phi_phi_p}) gives the number of walls. If $w$ and $w^{\prime}$ are the winding numbers of fields $\phi$ and $\phi^{\prime}$, respectively, then the number of DWs bounded by Type-$\phi^{\prime}$ strings is given by~\cite{Rothstein:1992rh, Barr:1992qq} $N_W^{\phi^{\prime}} = |nw+n^{\prime}w^{\prime}|_{\text{min}}$,
where the minimum is w.r.t varying $w \in \mathbb{Z}$.
Also, the energy density stored in a majoron field around a string (known as gradient energy) is proportional to $(nw+n^{\prime}w^{\prime})^2$~\cite{Rothstein:1992rh,Niu:2023khv}. Minimal Type-$\phi^{\prime}$ strings, which are expected to be the most abundant, possess $w^{\prime} = \pm 1$ and $w$ is tuned to minimise the gradient energy, which also minimises $N_W^{\phi^{\prime}}$. When $\phi$ acquires a VEV, Type-$\phi$ strings form even where Type-$\phi^{\prime}$ strings do not exist, and they have $w^{\prime}=0$. Hence, the number of DWs bounded by Type-$\phi$ strings is given by $N_W^{\phi} = |nw|$.
For the minimal Type-$\phi$ strings, we have $w= \pm 1$.
The `domain wall problem' does not exist if any of the three conditions are satisfied~\cite{Rothstein:1992rh, Barr:1992qq}:
\begin{align}
    n=1 \,,  \nonumber \\ 
    n^{\prime}=1 \,, \nonumber \\
    \exists \, w,w^{\prime}  \, \ni \, |nw+n^{\prime}w^{\prime}|=1 \, \, \, \text{for} \, n, n^{\prime} \neq 1 \, .
    \label{eq:oneDW}
\end{align}
In this paper, we explore only the scenario corresponding to $n=1  \, (n^{\prime}=d-1)$.
We leave $N_W > 1$ DW formation scenarios for a future study, which will require the introduction of further bias term(s) to make the stable DW network dissipate energy and emit GW. The details of the benchmark parameters, used to describe the scenarios, are summarized in Table~\ref{tab:BP} and their phenomenological aspects are outlined later in Table~\ref{tab:BP_DM}.

\begin{table}[h!]
\centering
\renewcommand{\arraystretch}{1.6}
\setlength{\tabcolsep}{7pt}

\begin{tabular}{|c|c|c|c||c|c|c||c|c|c|}
\hline
\multirow{2}{*}{Benchmarks} 
& \multirow{2}{*}{$d$} 
& \multirow{2}{*}{$n$} 
& \multirow{2}{*}{$n'$}
& \multicolumn{3}{c||}{Type-$\phi$ strings}
& \multicolumn{3}{c|}{Type-$\phi^\prime$ strings} \\
\cline{5-10}
& & & 
& $w$ & $w^\prime$ & $N_W^\phi = |nw|$
& $w$ & $w^\prime$ & $N_W^{\phi^\prime} = |nw+n'w'|_\text{min}$ \\
\hline
\hline

BP1 & 16 & 1 & 15
& 1 & 0 & 1
& -15 & 1 & 0 \\

\multicolumn{10}{c}{}\\[-23.2pt]
\hline
BP2 & 17 & 1 & 16
& 1 & 0 & 1
& -16 & 1 & 0 \\
\hline

\end{tabular}
\caption{Benchmark points (motivated by the relic abundance of the dark matter) showing the values of $d, n, n^\prime$ and the wingding numbers $(w,w^\prime)$ for both Type-$\phi$ and Type-$\phi^\prime$ strings, together with the corresponding numbers of domain walls attached to each type. The table highlights only the cases in which $N_W^\phi = 1$ and $N_W^{\phi'} = 0$.}

\label{tab:BP}
\end{table}


A CS is a topologically stable object, and a conserved topological flux flows along the string. Since Type-$\phi$ and Type-$\phi^{\prime}$ strings have different winding numbers, they form separate string loops, and mixed string loops do not form because mixed string loops do not conserve topological flux. In other words, there cannot be any string loop where one segment is made of a Type-$\phi$ string and another of a Type-$\phi^{\prime}$ string. Also, we assume Type-$\phi$ and Type-$\phi^{\prime}$ strings' evolution to proceed independently into distinct scaling regimes, with no interaction between them. Hence, Type-$\phi$ and Type-$\phi^{\prime}$ string loops vibrate with separate string tensions and radiate GW.

Type-$\phi^{\prime}$ strings are formed in this model due to the breaking of the local $U(1)_{B-L}$ symmetry only. Thus, their string tension can still be expressed using Eq.~(\ref{eq.22}) and their GW spectrum is given by 
\begin{align}
    \Omega_{\text GW}(f) h^2 \Bigg|_{\phi^{\prime}} = \Omega_{\text{GW}}(f) h^2 \Bigg|_\text{LCS}= h^2 \sum_k \Omega_{\text{GW}}^{(k)}(f)\Bigg|_\text{LCS} \, ,
\end{align}
where the expression for $\Omega_{\text{GW}}^{(k)}(f)$ is provided in Eq.~(\ref{eq.24}). 
Consequently, for Type-$\phi^{\prime}$ strings we still use the four simulation templates (\texttt{stable-k}, \texttt{stable-m}, \texttt{stable-n} and \texttt{stable-c}) given for GW spectra for LCS in the \texttt{PTArcade} package, instead of integrating Eq.~(\ref{eq.24}).

However, the majoron in our model is a pNGB and not massless. Hence, the results for GCS presented in Subsection~\ref{gcs} are not applicable for Type-$\phi$ strings here. So, we expect the majoron mass to have some influence on the GW spectrum. In the literature, simulation result exists for the GW spectrum from CSs resulting from the breaking of an approximate global $U(1)$ symmetry, which also gives rise to a pNGB candidate in the early Universe. These simulations have been performed in the context of axions~\cite{Gorghetto:2021fsn, Gorghetto:2018myk}. In such a scenario, the string tension of CSs in the scaling regime is given by 
\begin{align}
\mu_{\phi} = \pi \eta^2 \log \left( \frac{m_{\rho} \alpha^{\prime}}{H \sqrt{\xi}} \right)
 \,,
 \label{eq.mu_phiprime}
\end{align}
where $m_{\rho}=(\sqrt{2\lambda})\eta$ is the mass of the heavy radial field $\rho$ and $\alpha^\prime$ is a dimensionless parameter that characterises the archetypal shape of the CSs in the scaling regime. The parameter $\xi$ has already been introduced in Subsection~\ref{gcs}. 

The GW spectrum from CSs arising via a softly broken accidental global $U(1)$ symmetry, as simulated in Ref.~\cite{Gorghetto:2021fsn}, is well-approximated by the analytical expression
\begin{align}
\Omega_{\text GW}(f) h^2 \Bigg|_{\phi} = 0.80 \times 10^{-15} \left( \frac{\eta}{10^{14} \, \text{GeV}} \right)^4 \left( \frac{10}{g_f} \right)^{\frac{1}{3}} \left\{ 1 + 0.12 \log \left[ \left( \frac{m_{\rho}}{10^{14} \, \text{GeV}} \right) \left( \frac{10^{-8} \, \text{Hz}}{f} \right)^2 \right] \right\}^4
 \,.
 \label{gw_phi}
\end{align}
Here, $g_f$ stands for the effective number of degrees of freedom that are in thermal equilibrium at the temperature $T_f$, where $T_f$ is the temperature of the Universe when GWs of a particular frequency $f$ are dominantly produced. The simulation results of Ref.~\cite{Gorghetto:2021fsn} suggest that the pNGB mass, $m_{\chi}$, regulates only the position of the infrared (IR) cut-off of the above spectrum, which goes as $\sqrt{m_{\chi}}$. $m_{\chi}$ does not affect the GW emitted during the scaling regime, as it only determines when the string network gets destroyed. Hence, the above GW spectrum is valid all the way down to the nano-Hz frequency range ($\sim 10^{-9}$ Hz) only if $m_{\chi}\lesssim 10^{-15}$ eV. Otherwise, the $m_{\chi} = 3 H(T_{*})$ condition will be satisfied at a higher temperature, and the GW spectrum will hit the IR cut-off and fall sharply before the nano-Hz frequency range.

Similarly the $m_{\chi} = 3 H(T_{*})$ condition has some impact on Type-\( \phi' \) LCSs as well, as at temperature $T_{*}$ the whole string-wall network collapses. Hence, the string-wall network should survive to such a temperature $T_{*}$ so that the GW spectrum of LCSs should extend down to $f \sim 10^{-9}$ Hz. The frequency-temperature relationship for LCSs can be found in Eq.~(3.5) of Ref.~\cite{Cui:2018rwi}, and the relation has a VEV dependence. Using this equation and $m_{\chi} = 3 H(T_{*})$ we find that for $\eta^{\prime} \sim 10^{14} \, (10^{13})$ GeV, the Type-\( \phi' \) LCS GW spectrum spans the NANOGrav frequency range for $m_{\chi} \lesssim 10^{-23} \, (10^{-25})$ eV. We get different cut-off values of $m_{\chi}$ for the LCS and GCS-like spectra of Eq.~(\ref{gw_phi}) because the $f-T$ relationships are different for them~\cite{Cui:2018rwi, Chang:2021afa}.

Finally, as discussed earlier, Type-$\phi $ and Type-$\phi^{\prime}$ strings possess distinct winding numbers and string tensions. Thus, we assume they evolve independently into their respective scaling regimes without influencing each other. Hence, the total GW relic is the sum of GW contributions coming from Type-$\phi$ and Type-$\phi^{\prime}$ strings and can be expressed as
\begin{align}
{\Omega}^{\text{total}}_{\text{GW}}(f)= \Omega_{\text GW}(f) h^2 \Bigg|_{\phi^{\prime}} + \Omega_{\text GW}(f) h^2 \Bigg|_{\phi}
\,.
 \label{eq.total}
 \end{align}
In short, for the modified majoron model, we perform our analysis using four different \texttt{PTArcade} LCS templates for Type-$\phi^\prime$ strings, and for each of these templates, we consider Eq.~(\ref{gw_phi}) for the GW spectrum of Type-$\phi$ strings. We summarize these four scenarios in Table~\ref{t.model}. 

Another comment is in order for the GW spectrum of the modified majoron model. When the $m_{\chi} = 3 H(T_{*})$ condition is achieved, DWs will form and assuming that one of the conditions of Eq.~(\ref{eq:oneDW}) is satisfied, the system of DW and strings will collapse. This process will emit additional GWs supplementing that from Eq.~(\ref{eq.total}). However, the spectrum of GWs from the collapse of the Type-$\phi$ CS-DW system is not well known but is expected to have amplitudes and frequencies of the same order as the last e-folding of the scaling regime~\cite{Hiramatsu:2012sc}. Hence, the IR cut-off behaviour of the GW spectrum is expected to be affected. However, if the majoron mass is significantly lower than $10^{-23}$ eV, this contribution is unimportant for our study as the IR cut-off will safely fall below the nano-Hz range. On the other hand, the effect of DWs destroying LCSs is known, but for $T_{*} \sim 10^{-7}$ GeV corresponding to $m_{\chi} \sim 10^{-23}$ eV, the deviation from a pure LCS GW spectrum is expected to be minimal~\cite{Dunsky:2021tih}.

Hence, for $m_{\chi} < 10^{-23}$ eV, we expect the full GW spectrum given by Eq.~(\ref{eq.total}) to contribute to the stochastic GW background probed by NANOGrav. However, for $10^{-23} \, \text{eV} <m_{\chi} < 10^{-15} \, \text{eV}$, we expect the GW spectrum from Type-$\phi^{\prime}$ LCS to die down before $10^{-8}$ Hz but the GW spectrum from Type-$\phi$ strings still reaches below nano-Hz. We shall consider both scenarios in the next section.

\begin{table}[!tp]
\centering
\renewcommand{\arraystretch}{1.6} 
\setlength{\tabcolsep}{0pt} 
\setlength{\arrayrulewidth}{0.4pt} 
\arrayrulecolor{black} 
\begin{tabular}{|c|c|c|}
\hline
\multicolumn{1}{|c|}{\hspace{4.5pt}Total\hspace{4.5pt}} & Type-$\phi$ & Type-$\phi^\prime$ \\ \hline \hline
\hspace{4.5pt}Type-A\hspace{4.5pt} &  \multirow{4}{*}{\hspace{4.5pt}Eq.~(\ref{gw_phi})\hspace{4.5pt}} & \hspace{4.5pt}\texttt{stable-c}\hspace{4.5pt} \\ 
\hspace{4.5pt}Type-B\hspace{4.5pt} &  & \hspace{4.5pt}\texttt{stable-k}\hspace{4.5pt} \\ 
\hspace{4.5pt}Type-C\hspace{4.5pt} & & \hspace{4.5pt}\texttt{stable-m}\hspace{4.5pt} \\ 
\hspace{4.5pt}Type-D\hspace{4.5pt} & & \hspace{4.5pt}\texttt{stable-n}\hspace{4.5pt} \\ \hline
\end{tabular}
\caption{Four different cases for the total GW spectrum from CSs formed in the modified majoron model}  
\label{t.model}
\end{table}

\section{Results}
\label{sec:results}

In this section, we compare the evidence of the NANOGrav~\cite{NANOGrav:2023gor, NANOGrav:2023hde} stochastic GW signal with GW spectra obtained from GCSs (within the simplest majoron model) and the CSs that are formed in the modified majoron model. 
Additionally, we demonstrate how these results can be utilised to constrain the parameters of the respective models.

Before describing our results, first, let us provide a brief outline of our method to obtain the best-fit parameters for the respective models. For this purpose, we use a Markov-Chain Monte Carlo (MCMC) Bayesian analysis by means of the \texttt{PTArcade}~\cite{Mitridate:2023oar} package in the \texttt{enterprise}~\cite{ellis_2020_4059815} mode. Let us assume that $\Vec{\theta}$ denotes the parameters that characterize the GW spectrum within a BSM model, and $\mathcal{D}$ represents the PTA data. If $\mathcal{P}(\Vec{\theta})$ and $\mathcal{P}(\mathcal{D})$ represent the distribution of the priors of the parameters and the marginalized likelihood, respectively, then the posterior distribution $\mathcal{P}(\Vec{\theta}|\mathcal{D})$ of model parameters is given by
\begin{align}
\mathcal{P}(\Vec{\theta}|\mathcal{D}) = \frac{\mathcal{P}(\mathcal{D}|\Vec{\theta}) \mathcal{P}(\Vec{\theta})}{\mathcal{P}(\mathcal{D})} \, .
\label{eq:posterior}
\end{align}
Here, $\mathcal{P}(\mathcal{D}|\Vec{\theta})$ denotes the likelihood for the PTA dataset.

One of the advantages of using Bayesian analysis is that one can also use it to readily compute the Bayes factor (BF) in favour of a model $Y$ against a reference model $X$. Such a BF is given by
\begin{align}
\text{BF} \equiv \frac{\mathcal{P}(\mathcal{D}|Y)}{\mathcal{P}(\mathcal{D}|X)} \, .
\label{eq:BF}
\end{align}
From the astrophysical point of view, a population of SMBHB mergers is expected to be the primary source of stochastic GW in the PTA frequency range. Hence, we choose GWs from SMBHB mergers as the reference model $X$. We set the switches \texttt{smbhb=True} and \texttt{bhb\_th\_prior=True} in \texttt{PTArcade} to evaluate the BFs required for this analysis. \texttt{PTArcade} uses a bi-variate Gaussian prior for the amplitude ($A$) and spectral index ($\gamma$) for the GW spectrum from SMBHB mergers, which is discussed in Appendix~\ref{sec:smbhb}. 

Now, let us discuss the results obtained from fitting the NANOGrav 15-year data. First, we discuss the simplest majoron model followed by the modified majoron model.

\begin{figure}[tp!]
    \centering
        \includegraphics[width=0.5\linewidth]{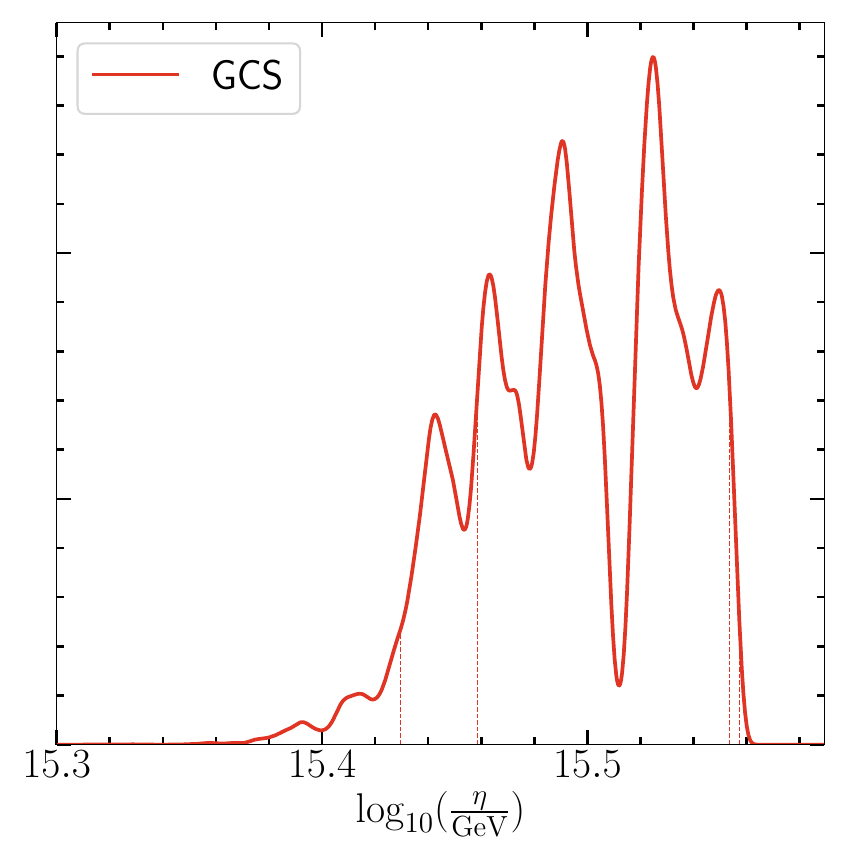}
    
  \caption{Posterior distributions of $\log_{10} \big(\frac{\eta}{\text{GeV}} \big)$ from comparing the simplified majoron model with the NANOGrav 15-year data when GCS is considered as the only source of stochastic GW background. Here, $\eta$ is the VEV of the complex singlet $\phi$ breaking the global $U(1)_L$ symmetry. Among the four red dashed lines the inner two depict the $1\sigma$ credible interval of $\log_{10}(\frac{\eta}{\textrm{GeV}})$ while the outer two lines are for $2\sigma$.
  }
    \label{fig:GCS_posterior}
\end{figure}
\begin{figure}[tp!]
    \centering    
        \includegraphics[width=0.7\linewidth]{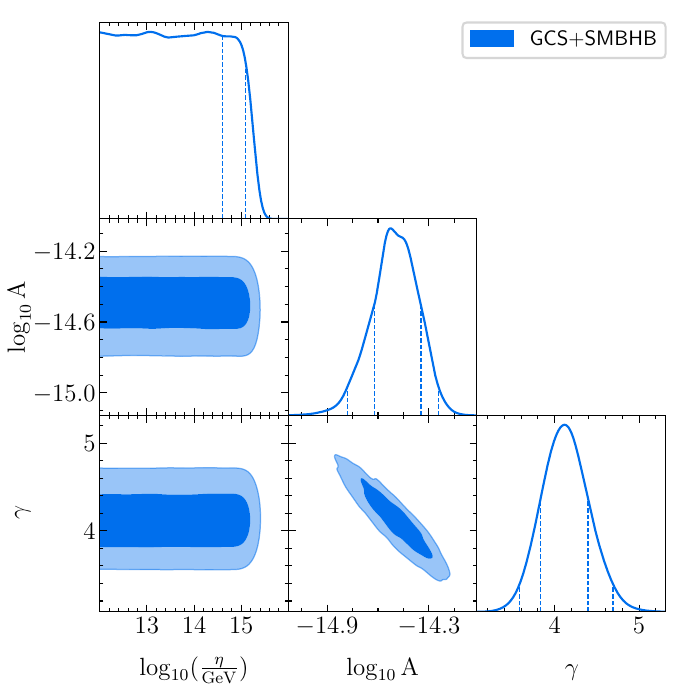}
    
    \caption{Posterior distributions of GCS and SMBHB parameters from comparing the simplified majoron model with the NANOGrav 15-year data in the GCS+SMBHB scenario. The darker (lighter) shade of blue signifies the $1\sigma$ ($2\sigma$) credible contours. Among the four dashed lines of both $\gamma$ as well as $\log_{10}$A the outer (inner) two describe the $2\sigma$ ($1\sigma$) credible intervals. However, for the $\log_{10}(\frac{\eta}{\textrm{GeV}})$ the left (right) one among the two dashed lines signify the upper limits of the $1\sigma$ ($2\sigma$) credible intervals.}
    \label{fig:GCS+SMBHB_posterior}
\end{figure}


\paragraph{The simplest majoron model:} The reader should recall that the simplest majoron model consists of an additional global $U(1)_{L}$ symmetry, which is spontaneously broken by a complex singlet scalar $\phi$ when it acquires a VEV, $\eta$. The spontaneous breaking of the global $U(1)_{L}$ symmetry leads to the formation of GCSs. The GW spectrum from such global strings is discussed in Section~\ref{gcs}. For the nano-HZ frequency range, the GW spectrum is essentially given in Eq.~(\ref{eq.21}). We compare this spectrum against the NANOGrav data here. One can notice from Eq.~(\ref{eq.21}) that the only BSM parameter that controls the GW spectrum is $\eta$. We incorporate this GW spectrum within \texttt{PTArcade} by using the \texttt{Model} file, and use a logarithmically (with base 10) uniform prior for $\eta$ from $10^{12}$ GeV up to $10^{16}$ GeV. 

In Fig.~\ref{fig:GCS_posterior}, we present the posterior distributions of the BSM parameter $\log_{10} \big(\frac{\eta}{\text{GeV}} \big)$, when the GW emission from GCSs is considered the only source of stochastic GW background. The fit to the NANOGrav 15-year data yields a narrow $2 \sigma$ Credible Interval of $2.69\times 10^{15} \, \text{GeV} < \eta < 3.63 \times 10^{15} \, \text{GeV}$.  However, when this fit is compared against the reference SMBHB merger GW spectrum, we obtain a tiny BF of $7.66 \times 10^{-5}$. That means that the GW emission from the SMBHB merger is obviously a much-preferred solution to the latest NANOGrav data compared to the GCS solution. We can surmise that the result is quite expected because of the negative ($f^{-1/3}$) power law of the GCS GW-spectrum, while NANOGrav data prefers a positive power law. Nevertheless, the above fit can be treated as a constraint on the global $U(1)_L$ breaking scale $\eta$ if one assumes GCSs as the only source of stochastic GW. It is also important to recall here that $\eta$ in the simplest majoron model shows up in the expression for light neutrino masses (cf. Eq.~(\ref{eq.3})). Cosmology places a strong bound on the sum of neutrino masses $\sum m_{\nu} \lesssim 0.1$ eV~\cite{ParticleDataGroup:2024cfk}. Hence, assuming the normal hierarchy (NH) of light neutrino masses and $m_1\approx m_2 \ll m_3 \sim 0.1$ eV, one can translate the bound on $\eta$ to a simplistic qualitative bound on the effective coupling $12.57 < \delta_{33} < 16.97$ using Eq.~(\ref{eq.3}). As for the inverted hierarchy of neutrino masses, since two of the heavier SM neutrinos are expected to be nearly degenerate, we refrain from deriving such simplistic bounds on any $\delta_{IJ}$ couplings. These results are summarized in Table~\ref{tab:gcs}.

\begin{table}[!tp]
\centering

\renewcommand{\arraystretch}{1.6} 

\setlength{\tabcolsep}{0pt} 

\setlength{\arrayrulewidth}{0.4pt} 

\arrayrulecolor{black} 

\begin{tabular}{|c|c|c|c|c|c|}
\hline
 GW & \multirow{2}{*}{\hspace{4.5pt}Parameters\hspace{4.5pt}} & \multirow{2}{*}{Priors} & \hspace{4.5pt}$2 \sigma$ Credible Intervals\hspace{4.5pt} & \hspace{4.5pt}$2 \sigma$ Credible Interval\hspace{4.5pt}  & \multirow{2}{*}{BF} \\
 spectrum & & &  of parameters & of $\delta_{33}$ & \\
 \hline
 \hline
 GCS & $\log_{10} \big(\frac{\eta}{\text{GeV}} \big)$ & [12,16] & [15.43, 15.56] & [12.57, 16.97] &  \hspace{4.5pt}$7.66 \times 10^{-5} \pm 0.0$\hspace{4.5pt} \\
 \hline
 \multirow{3}{*}{\hspace{4.5pt}GCS+SMBHB\hspace{4.5pt}} & $\log_{10} \big(\frac{\eta}{\text{GeV}} \big)$ & [12,16] & $ < 15.10$ & \multirow{3}{*}{ $<5.89$} & \multirow{3}{*}{$0.65 \pm 0.16$}\\

& $\log_{10} A$ & \multirow{2}{*}{\hspace{4.5pt}Eq.~(\ref{eq:SMBHB_prior})\hspace{4.5pt}} & [-14.80, -14.26] & & \\
& $\gamma$ & & [3.60, 4.70] & & \\
\hline
\end{tabular}
\caption{Parameters for the stochastic GW spectrum in the simplified majoron model under GCS-only and GCS+SMBHB scenarios. Shown are priors, $2\sigma$ Credible Intervals from fits to NANOGrav 15-year data, and BFs relative to the SMBHB-only case.  Also included are $2\sigma$ Credible Intervals on the coupling $\delta_{33}$, assuming $m_3 \sim 0.1$ eV and NH of neutrino masses (see text for more details). For the GCS+SMBHB case, we show only the upper limit of Credible Interval on $\eta$ and $\delta_{33}$ for reasons described in the text.
}  
\label{tab:gcs}
\end{table}

We redo the analysis again by taking the GCS together with SMBHB mergers as two sources of stochastic GW emission. The posterior distribution of $\log_{10} \big(\frac{\eta}{\text{GeV}} \big)$ for this analysis is shown in Fig.~\ref{fig:GCS+SMBHB_posterior} along with the same for the amplitude ($A$) and spectral index ($\gamma$) of the GW spectrum from SMBHB mergers. The posterior of $\log_{10} \big(\frac{\eta}{\text{GeV}} \big)$ remains flat below $\sim 10^{15}$ GeV. This behaviour is expected as the amplitude of the GCS GW spectrum decreases rapidly with decreasing $\eta$, and the resultant GW spectrum essentially becomes the SMBHB merger spectrum. We further check the consistency of this fact by lowering the prior of $\eta$ down to $10^6$ GeV, and the blue line still remains flat down to that value. Thus, the lower bound on $\eta$ is irrelevant in this scenario. The fit to the NANOGrav data provides a $2 \sigma$ upper bound of $\eta < 1.26 \times 10^{15}$ GeV. This result is tabulated again in Table~\ref{tab:gcs} together with the  $2\sigma$ Credible Intervals of $\log_{10}A$ and $\gamma$. 
Again, a comparison with the SMBHB-only solution leads to a BF of $0.65 \pm 0.16$. Hence, the SMBHB merger is still a slightly better fit for the NANOgrav 15-year data. Using the same assumptions regarding light neutrino mass spectrum as outlined in the above paragraph, we obtain a bound on the coupling $\delta_{33} < 5.89$.

\begin{figure}[!tp]
    \centering
    
    \subfigure{\includegraphics[width=0.495\textwidth]{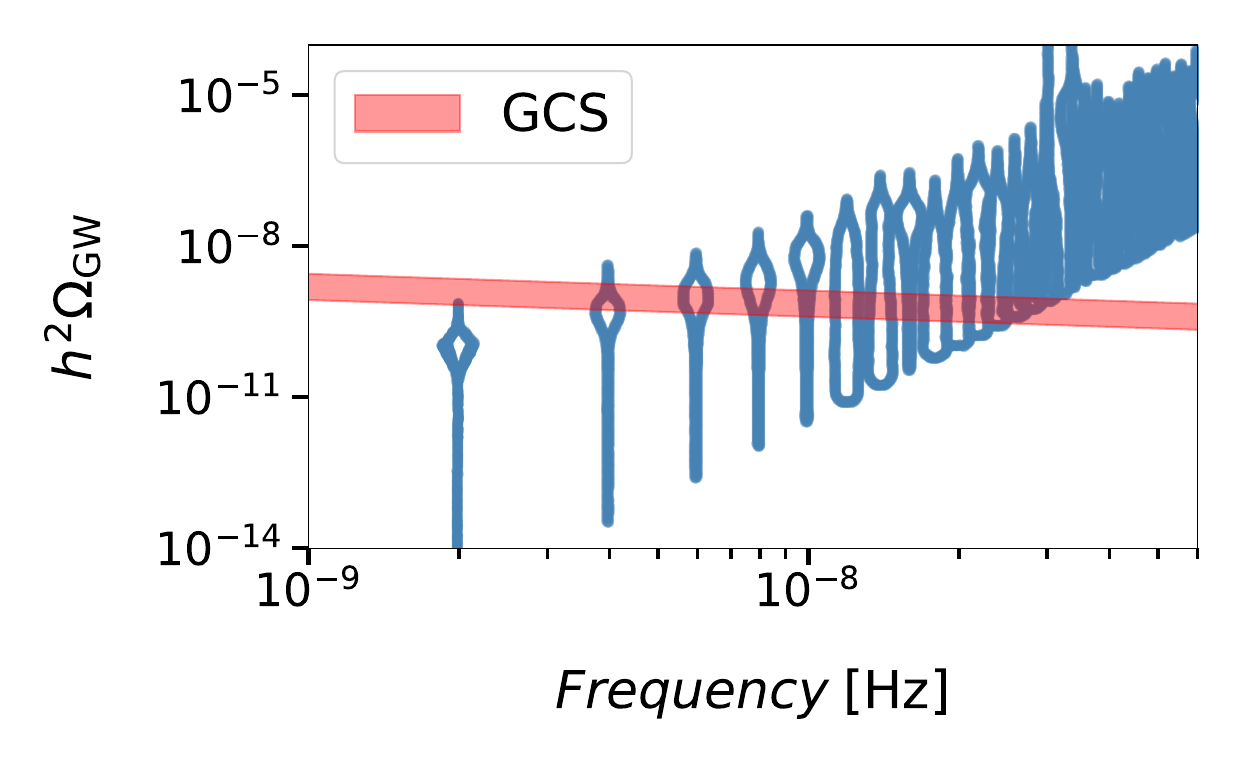}} 
    \subfigure{\includegraphics[width=0.495\textwidth]{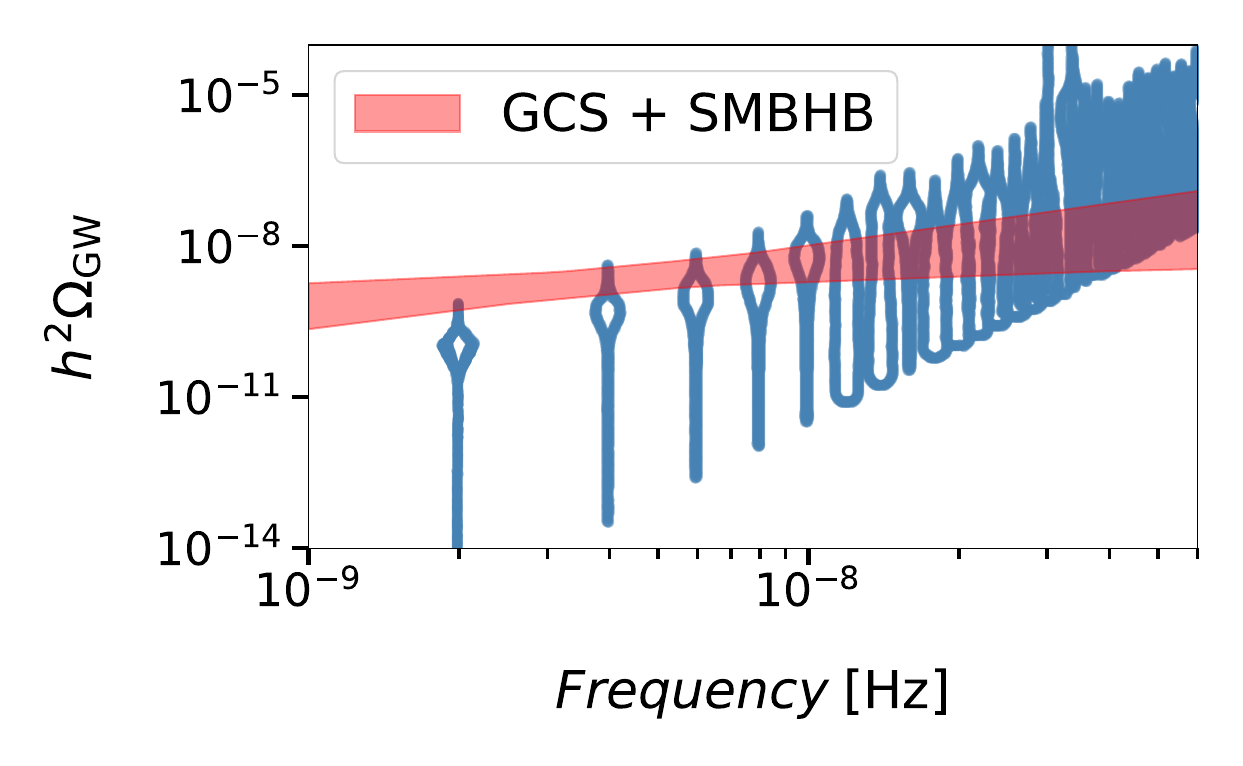}} \\
    
    \caption{The red bands show the GW spectrum for the GCS-only (left) and GCS+SMBHB (right) cases, obtained by varying model parameters within their $2\sigma$ Credible Intervals in the simplified majoron model. Blue violins represent the NANOGrav 15-year data.}
   \label{fig:GCS_NG15}
\end{figure}

Figure~\ref{fig:GCS_NG15} illustrates the predicted GW spectra for two scenarios: GCS-only (left panel) and GCS combined with the SMBHB contribution (right panel), shown by the red bands. We vary the model parameters within their $2\sigma$ Credible Intervals to obtain the red bands. The blue violin plots represent the NANOGrav 15-year data, reflecting the posterior distribution of the observed stochastic GW background.

\begin{table}[!tp]
\centering

\renewcommand{\arraystretch}{1.6} 

\setlength{\tabcolsep}{0pt} 

\setlength{\arrayrulewidth}{0.4pt} 

\arrayrulecolor{black} 

\begin{tabular}{|c|c|c|c|c|c|}
\hline
 GW & \multirow{2}{*}{\hspace{4.5pt}Parameters\hspace{4.5pt}} & \multirow{2}{*}{Priors} & \hspace{4.5pt}$2 \sigma$ Credible Intervals\hspace{4.5pt} & \hspace{4.5pt}$2 \sigma$ Credible Interval\hspace{4.5pt}  & \multirow{2}{*}{BF} \\
 spectrum & & &  of parameters & of $\delta_{33}$ & \\
 \hline
 \hline
 \multirow{2}{*}{\hspace{4.5pt}Type-A\hspace{4.5pt}} & $\log_{10} \big(\frac{\eta^{\prime}}{\text{GeV}} \big)$ & [11.8,16] & [13.67, 14.00] & \multirow{2}{*}{ $<0.15$} & \multirow{2}{*}{\hspace{4.5pt}$0.07 \pm 0.01$\hspace{4.5pt}}\\
 & $\log_{10} \big(\frac{\eta}{\text{GeV}} \big)$ & [6,16] & $<13.50$ & & \\
 \hline
 \multirow{4}{*}{\hspace{4.5pt}Type-A+SMBHB\hspace{4.5pt}} & $\log_{10} \big(\frac{\eta^{\prime}}{\text{GeV}} \big)$ & [11.8,16] &   [12.46, 13.93] & \multirow{4}{*}{ $<0.03$} & \multirow{4}{*}{\hspace{4.5pt}$0.28 \pm 0.05$\hspace{4.5pt}}\\
& $\log_{10} \big(\frac{\eta}{\text{GeV}} \big)$ & [6,16] & $ < 12.85$ & & \\
& $\log_{10} A$ & \multirow{2}{*}{\hspace{4.5pt}Eq.~(\ref{eq:SMBHB_prior})\hspace{4.5pt}} & [-16.30, -14.20] &  & \\
& $\gamma$ & & [3.60, 5.20] & & \\
\hline
\hline
 \multirow{2}{*}{\hspace{4.5pt}Type-B\hspace{4.5pt}} & $\log_{10} \big(\frac{\eta^{\prime}}{\text{GeV}} \big)$ & [11.8,16] & [13.58, 13.87] & \multirow{2}{*}{ $<0.15$} & \multirow{2}{*}{\hspace{4.5pt}$0.15 \pm 0.02$\hspace{4.5pt}}\\
 & $\log_{10} \big(\frac{\eta}{\text{GeV}} \big)$ & [6,16] & $<13.50$ & & \\
 \hline
 \multirow{4}{*}{\hspace{4.5pt}Type-B+SMBHB\hspace{4.5pt}} & $\log_{10} \big(\frac{\eta^{\prime}}{\text{GeV}} \big)$ & [11.8,16] & $ [12.12,13.84]$ & \multirow{4}{*}{ $<0.03$} & \multirow{4}{*}{\hspace{4.5pt}$0.35 \pm 0.05$\hspace{4.5pt}}\\
& $\log_{10} \big(\frac{\eta}{\text{GeV}} \big)$ & [6,16] & $ < 12.85$ & & \\
& $\log_{10} A$ & \multirow{2}{*}{\hspace{4.5pt}Eq.~(\ref{eq:SMBHB_prior})\hspace{4.5pt}} & [-16.20, -14.30] &  & \\
& $\gamma$ & & [3.70, 5.10] & & \\
\hline
\hline
 \multirow{2}{*}{\hspace{4.5pt}Type-C\hspace{4.5pt}} & $\log_{10} \big(\frac{\eta^{\prime}}{\text{GeV}} \big)$ & [11.8,16] & [13.47, 13.78] & \multirow{2}{*}{ $<0.09$} & \multirow{2}{*}{\hspace{4.5pt}$0.15 \pm 0.02$\hspace{4.5pt}}\\
 & $\log_{10} \big(\frac{\eta}{\text{GeV}} \big)$ & [6,16] & $<13.30$ & & \\
 \hline
 \multirow{4}{*}{\hspace{4.5pt}Type-C+SMBHB\hspace{4.5pt}} & $\log_{10} \big(\frac{\eta^{\prime}}{\text{GeV}} \big)$ & [11.8,16] & $ [12.10,13.74]$ & \multirow{4}{*}{ $<0.03$} & \multirow{4}{*}{\hspace{4.5pt}$0.28 \pm 0.04$\hspace{4.5pt}}\\
& $\log_{10} \big(\frac{\eta}{\text{GeV}} \big)$ & [6,16] & $ < 12.80$ & & \\
& $\log_{10} A$ & \multirow{2}{*}{\hspace{4.5pt}Eq.~(\ref{eq:SMBHB_prior})\hspace{4.5pt}} & [-16.20, -14.40] &  & \\
& $\gamma$ & & [3.60, 5.20] & & \\
\hline
\hline
 \multirow{2}{*}{\hspace{4.5pt}Type-D\hspace{4.5pt}} & $\log_{10} \big(\frac{\eta^{\prime}}{\text{GeV}} \big)$ & [11.8,16] & [13.65, 13.95] & \multirow{2}{*}{ $<0.15$} & \multirow{2}{*}{\hspace{4.5pt}$0.12 \pm 0.02$\hspace{4.5pt}}\\
 & $\log_{10} \big(\frac{\eta}{\text{GeV}} \big)$ & [6,16] & $<13.50$ & & \\
 \hline
 \multirow{4}{*}{\hspace{4.5pt}Type-D+SMBHB\hspace{4.5pt}} & $\log_{10} \big(\frac{\eta^{\prime}}{\text{GeV}} \big)$ & [11.8,16] & [12.50, 13.88] & \multirow{4}{*}{ $<0.04$} & \multirow{4}{*}{\hspace{4.5pt}$0.42 \pm 0.07$\hspace{4.5pt}}\\
& $\log_{10} \big(\frac{\eta}{\text{GeV}} \big)$ & [6,16] & $ < 12.95$ & & \\
& $\log_{10} A$ & \multirow{2}{*}{\hspace{4.5pt}Eq.~(\ref{eq:SMBHB_prior})\hspace{4.5pt}} & [-16.30, -14.20] &  & \\
& $\gamma$ & & [3.60, 5.20] & & \\
\hline
\end{tabular}
\caption{Parameters for the stochastic GW spectrum in the modified majoron model under CS-only and CS+SMBHB scenarios for $m_{\chi} < 10^{-23}$ eV. Shown are priors, $2\sigma$ Credible Intervals from fits to NANOGrav 15-year data, and BFs relative to the SMBHB-only case. Also included are upper limit of the $2\sigma$ Credible Intervals on the coupling $\delta_{33}$, assuming $m_3 \sim 0.1$ eV and NH of neutrino masses (see text for more details). We show only the upper limit of Credible Interval on $\eta$ and $\delta_{33}$ for reasons described in the text.
}  
\label{tab:cs_mmm}
\end{table}

\paragraph{The modified majoron model:} In the modified majoron model, the SM gauge group is extended by a local $U(1)_{B-L}$ and an approximate global $U(1)$ symmetry. When $\phi^{\prime}$ acquires a VEV $(\eta^{\prime})$, it spontaneously breaks $U(1)_{B-L}$. If $\eta^{\prime} > \eta$, the global $U(1)$ remains intact, up to a small soft breaking induced by the operator in Eq.~(\ref{eq.4}). Later, when $\phi$ develops a VEV $(\eta)$, the global $U(1)$ is also spontaneously broken. We assume that one of the conditions of Eq.~(\ref{eq:oneDW}) is satisfied so that only one DW is attached to either Type-$\phi$ or Type-$\phi^{\prime}$ strings. Then, in the scaling regime, the stochastic GW spectrum for this model is given by Eq.~(\ref{eq.total}) for $m_{\chi} < 10^{-23}$ eV. The spectrum is a combination of the GW spectrum from LCS (Type-$\phi^{\prime}$ strings) and the GCS-like spectrum (Type-$\phi$ strings), with an IR cutoff regulated by the majoron mass. We consider four different cases, Type-A, Type-B, Type-C, and Type-D, for the GW spectra from this model, depending on the LCS template as described in Table~\ref{t.model} (for more details, please see Sections~\ref {lcs} and~\ref {gcs+lcs}). In this paper, we focus exclusively on the $\eta < \eta^{\prime}$ scenario.
The BSM parameters that enter the GW spectrum in Eq.~(\ref{eq.total}) are VEVs $\eta^{\prime}$ and $\eta$. We adopt uniform priors for $\log_{10} \big(\frac{\eta^{\prime}}{\text{GeV}} \big)$ and $\log_{10} \big(\frac{\eta}{\text{GeV}} \big)$ in the range [11.8, 16] and [6, 16], respectively~\footnote{
We do not attempt to lower the prior range for the VEV $\eta^{\prime}$ for Type-$\phi^{\prime}$ LCS. This limitation stems from using the \texttt{PTArcade} LCS templates, which are valid in the range of $[-14, 0]$ for $\log_{10}(\text{G}\mu)$—equivalent to $\log_{10} \big(\frac{\eta^{\prime}}{\text{GeV}} \big)$ spanning from 11.83 to 18.83.
}. 

However, as discussed at the end of Section~\ref{gcs+lcs}, in the mass range $10^{-23} \, \text{eV} < m_{\chi} < 10^{-15} \, \text{eV}$, $m_{\chi}=3H(T_{*})$ will be satisfied at such values of $T_{*}$ that the GW spectrum from Type-$\phi^{\prime}$ LCS is expected to diminish significantly before reaching $10^{-8}$ Hz, whereas the spectrum from Type-$\phi$ strings continues to extend into the sub-nano-Hz range.

Let us first discuss the results for $m_{\chi} < 10^{-23}$ eV since, in this regime, the GW spectrum is quite different from the simplified majoron model. Table~\ref{tab:cs_mmm} presents the results of our fit of the GW spectra from CSs formed in the modified majoron model to the NANOGrav 15-year data. The table includes the fit results for all four cases of the GW spectra in this model, categorized as Type-A through Type-D, and provides a comparison between the CS-only and CS+SMBHB scenarios. A perceptive reader will notice that the results for GW spectra Types A through D are pretty similar across both scenarios. Consequently, we select the Type-C spectrum as a representative case for further analysis in this subsection, as this case provides slightly stronger limits. It is worth recalling that, in the Type-C case, we use the \texttt{PTArcade} template \texttt{stable-m} for the GW spectrum of Type-$\phi^{\prime}$ LCS, where the GW emission from the fundamental vibrational mode of closed loops is taken into account.

\begin{figure}[!tp]
    \centering
        \includegraphics[width=0.6\linewidth]{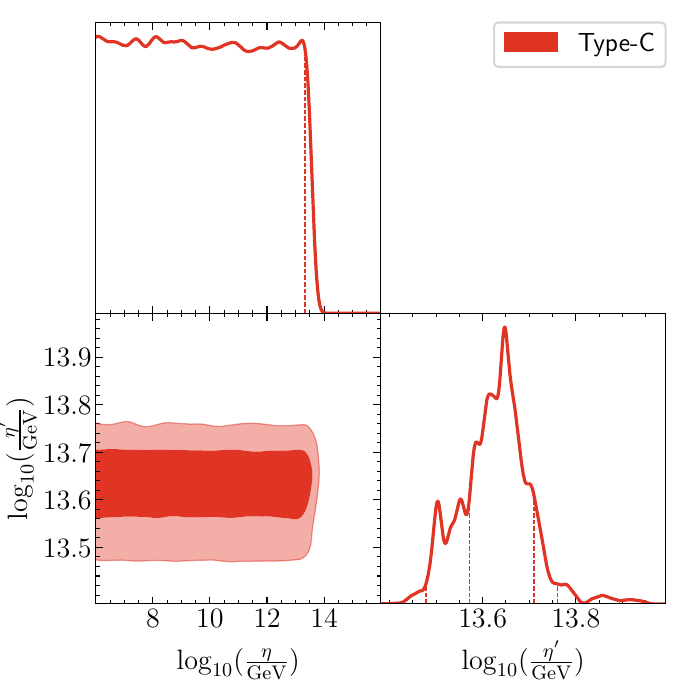}
    
    \caption{Posterior distributions of $\log_{10} \big(\frac{\eta^{\prime}}{\text{GeV}} \big)$ and $\log_{10} \big(\frac{\eta}{\text{GeV}} \big)$ from comparing the modified majoron model with the NANOGrav 15-year data when $m_{\chi} < 10^{-23}$ eV, and Type-$\phi$ and Type-$\phi^{\prime}$ CSs are considered as the only source of GW. For the model, we show the distributions for the Type-C GW spectrum only. Types A, B, and D GW spectra provide similar posterior distributions. Here, $\eta^{\prime} \, (\eta)$ is the VEV of the complex singlet $\phi^{\prime} \, (\phi)$ breaking the local $U(1)_{B-L}$ (accidental global $U(1)$) symmetry. The red dashed lines for $\log_{10}(\frac{\eta^\prime}{\textrm{GeV}})$ represent the different $1-2\sigma$ variable intervals in a similar manner as in Fig.~\ref{fig:GCS_posterior}. For $\log_{10}(\frac{\eta}{\textrm{GeV}})$ the $1\sigma$ and $2\sigma$ upper limits are superimposed on each other and are depicted using red dashed lines. Also, the contour shading follows the same convention as Fig.~\ref{fig:GCS+SMBHB_posterior}.
     }
    \label{fig:fig3}
\end{figure}

The posterior distributions of parameters $\log_{10} \big(\frac{\eta^{\prime}}{\text{GeV}} \big)$ and $\log_{10} \big(\frac{\eta}{\text{GeV}} \big)$ are presented in Fig.~\ref{fig:fig3} when CSs in the modified majoron model are considered as the only source of GW~\footnote{We should reiterate that Fig.~\ref{fig:fig3} contains the distributions from the Type-C GW spectrum only. Types A, B, and D GW spectra produce similar posterior distributions, and for brevity, we do not show them.}. Clearly, the above figure informs us that a narrow $2 \sigma$ Credible Interval of $2.95\times 10^{13} \, \text{GeV} < \eta^{\prime} < 6.03 \times 10^{13} \, \text{GeV}$ fits the NANOGrav data. In contrast, the posterior distribution of $\eta$ remains flat below $\sim 10^{13}$ GeV. Hence, the lower limit of the Credible Interval of $\eta$ is again irrelevant, and the fit yields a $2 \sigma$ upper limit of $\eta < 1.99 \times 10^{13}$ GeV. Hence, one can conclude that the NANOGrav data is predominantly fit by GWs emitted by Type-$\phi^{\prime}$ strings formed due to the breaking of the local $U(1)_{B-L}$ symmetry, and the Type-$\phi$ strings provide only a small component. Once again, a comparison with the SMBHB-only solution yields a BF of $0.15 \pm 0.02$, indicating that the SMBHB merger scenario still provides a better fit to the NANOGrav 15-year data. In the modified majoron model, the phase of $\phi$ acts as the majoron, and Eq.~(\ref{eq.3}) for light neutrino masses still holds. Thus, the VEV ($\eta$) acquired the $CP$-even component of $\phi$ will enter Eq.~(\ref{eq.3}), and using the same assumptions about the light neutrino mass spectrum as in the simplified majoron model, and following the same procedure, we derive a bound on the coupling: $\delta_{33} < 0.09$.

\begin{figure}[!tp]
    \centering
        \includegraphics[width=0.8\linewidth]{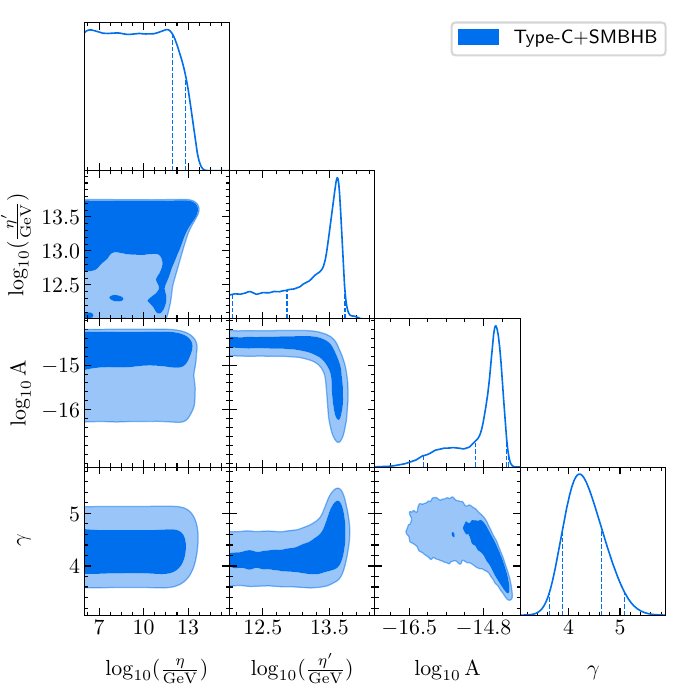}
    
    \caption{Posterior distributions of Type-$\phi$ and Type-$\phi^{\prime}$ CSs, and SMBHB parameters from comparing the modified majoron model with the NANOGrav 15-year data for $m_{\chi} < 10^{-23}$ eV in the CS+SMBHB scenario. We show the distributions for the Type-C GW spectrum only. Types A, B, and D GW spectra provide similar posterior distributions. The contour shading and dashed lines follow the same convention as in Fig.~\ref{fig:GCS+SMBHB_posterior}.}
    \label{fig:figx}
\end{figure}


For the CS+SMBHB scenario in the modified majoron model, the posterior distributions are shown in Fig.~\ref{fig:figx}. For the Type-C case in this scenario, a relatively broad $2 \sigma$ Credible Interval of $1.26 \times 10^{12} \, \text{GeV} < \eta^{\prime} < 5.50 \times 10^{13} \, \text{GeV}$ fits the data.
Again, the posterior distribution of $\eta$ remains flat below $\sim 10^{12}$ GeV, and the fit yields a $2 \sigma$ upper limit of $\eta < 6.31 \times 10^{12}$ GeV.
When this $2 \sigma$ upper limit is translated to a bound on the coupling entering the heaviest light neutrino mass, it provides $\delta_{33} < 0.03$. The BF in this case is $0.28 \pm 0.04$, which again indicates that the SMBHB-only scenario is a preferred solution.

However, there is a crucial difference between the CS+SMBHB scenario in the modified majoron model and the GCS+SMBHB scenario in the simplified majoron model. While the $\log_{10} \big(\frac{\eta}{\text{GeV}} \big)$ shows a flat posterior distribution for the latter case, the posterior of $\log_{10} \big(\frac{\eta^{\prime}}{\text{GeV}} \big)$ for Type-$\phi^{\prime}$ strings display a peak in the former case. Hence, we obtain a $2 \sigma$ lower limit for $\eta^{\prime}$ in the modified majoron model. This difference can be further reinforced by observing the $2 \sigma$ Credible Intervals in Table~\ref{tab:cs_mmm} for SMBHB parameters $\log_{10} A$ and $\gamma$ in the modified majoron model. A much larger range of these parameters is allowed in this model compared to the simplified majoron model. A careful look at the $\log_{10} A - \log_{10} \big(\frac{\eta^{\prime}}{\text{GeV}} \big)$ plane in Fig.~\ref{fig:figx} will provide an answer to this distinct behaviour. The figure shows two distinct arms for the fit to the NANOGrav data, with one arm corresponding to the peak in the $\log_{10} \big(\frac{\eta^{\prime}}{\text{GeV}} \big)$ posterior distribution. A similar but less prominent feature can be seen in the $\gamma - \log_{10} \big(\frac{\eta^{\prime}}{\text{GeV}} \big)$ plane also. This means that the data allows two different kinds of solutions. In one case,  for $\eta^{\prime} \lesssim 10^{13}$ GeV, SMBHB mergers provide the dominant GW spectrum and Type-$\phi^{\prime}$ LCSs is a secondary subdominant source, and this case provides the highest posterior value. In the other case, for $ 10^{13} \, \text{GeV} \lesssim \eta^{\prime} < 5.50 \times 10^{13} \, \text{GeV}$, Type-$\phi^{\prime}$ strings is the primary GW source, allowing for relaxed amplitude and much softer power law for the GW spectrum from SMBHB mergers. In both cases, Type-$\phi$ GCS-like strings provide only a tiny contribution to the total GW spectrum. In contrast, for the GCS+SMBHB scenario in the simplified majoron model, the data never allows a GCS-dominated parameter space, as discussed earlier.

\begin{figure}[!tp]
    \centering
    
    \subfigure{\includegraphics[width=0.495\textwidth]{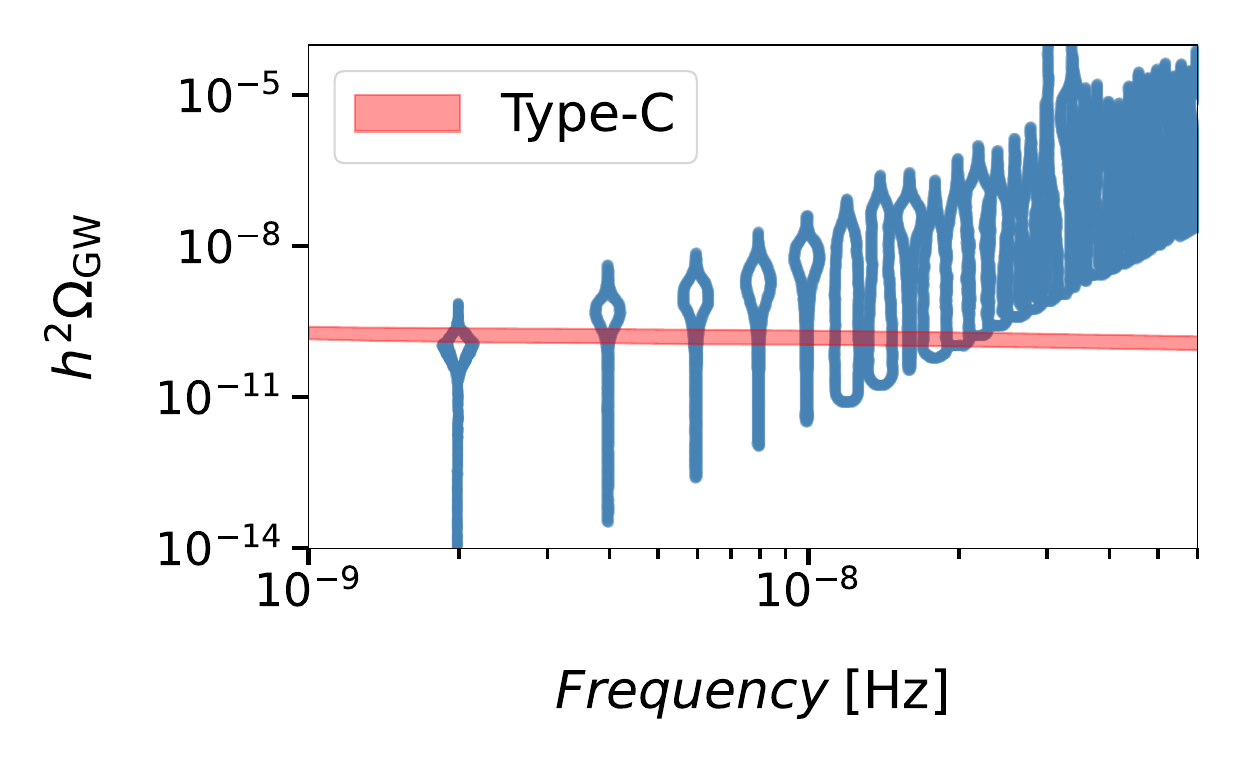}} 
    \subfigure{\includegraphics[width=0.495\textwidth]{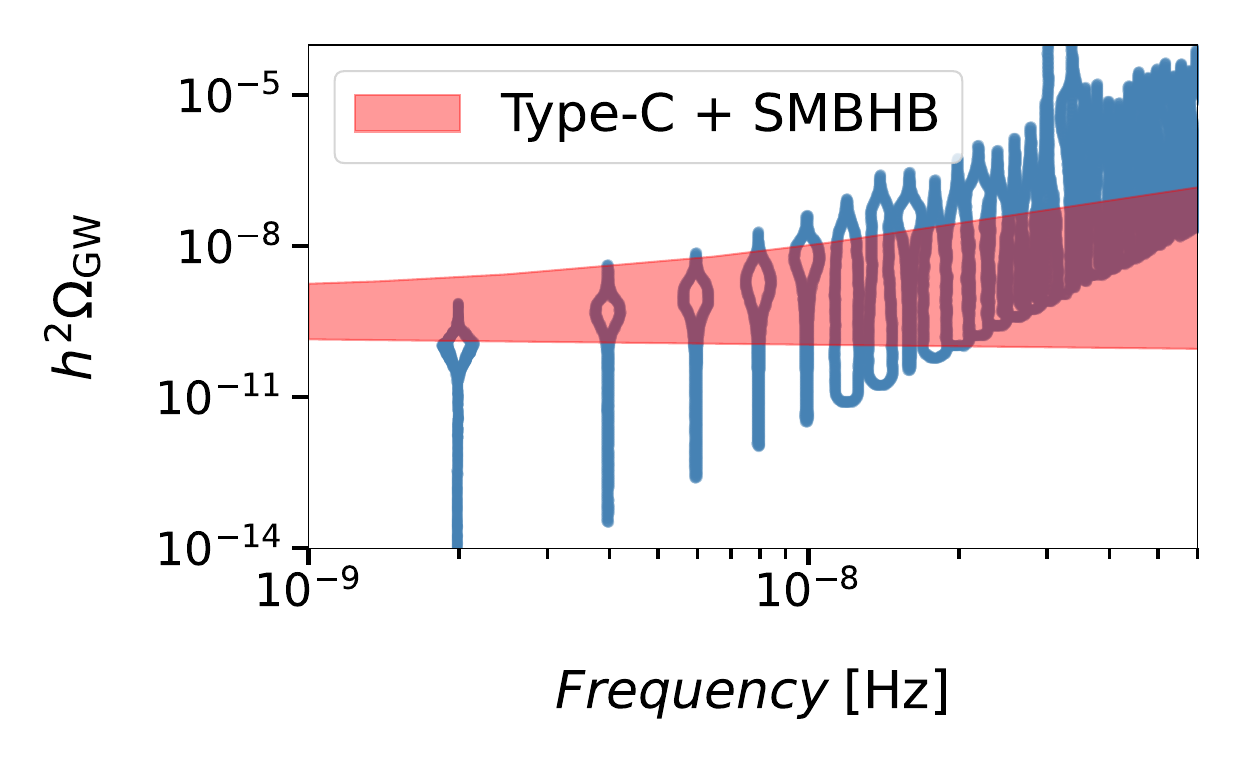}} \\
    
    \caption{The red bands show the GW spectra for the CS-only (left) and CS+SMBHB (right) cases, obtained by varying model parameters within their $2\sigma$ Credible Intervals in the modified majoron model for $m_{\chi} < 10^{-23}$ eV. Blue violins represent the NANOGrav 15-year data. For the model, we show the distributions for the Type-C GW spectrum only.}
   \label{fig:Type-C_NG15}
\end{figure}

Figure~\ref{fig:Type-C_NG15} shows the predicted gravitational wave spectra for two cases in the modified majoron model: CS-only (left panel) and CS with an added SMBHB contribution (right panel), indicated by red bands. These bands result from varying model parameters within their $2\sigma$ credible intervals. The blue violin plots depict the NANOGrav 15-year data, representing the posterior distribution of the observed stochastic GW background. For the model, we show the distributions for the Type-C GW case only, and the GW spectra are valid for $m_{\chi} < 10^{-23}$ eV.

Now, let us consider the mass range, $10^{-23} \, \text{eV} < m_{\chi} < 10^{-15} \, \text{eV}$ in the modified majoron model. As discussed earlier, in this regime the condition $m_{\chi} = 3H(T_{*})$ is met at values of $T_{*}$ where the GW spectrum from Type-$\phi^{\prime}$ LCS is expected to drop off well before $10^{-8}$ Hz, while the spectrum from Type-$\phi$ strings extends below the nano-Hz scale. Hence, the GW spectrum in this majoron mass range is given by the GCS-like GW spectrum of Eq.~(\ref{gw_phi}). The accidental $U(1)$ global symmetry breaking scale $\eta$ is the only BSM parameter, and we adopt a log-uniform (base 10) prior for $\eta$ ranging from $10^{12}$ GeV to $10^{16}$ GeV.

Since the GW spectrum of Eq.~(\ref{gw_phi}) is not significantly different from that of pure GCS, we expect results similar to those of the simplified majoron model. In the Type-$\phi$ strings only scenario, the fit to the NANOGrav data returns a $2\sigma$ Credible Interval of $2.75\times 10^{15} \, \text{GeV} < \eta < 3.63 \times 10^{15} \, \text{GeV}$, which is almost identical to the GCS-only scenario in the simplified majoron model. When Type-$\phi$ strings are considered as a secondary source of GW emission on top of SMBHB mergers, we obtain $2\sigma$ Credible Intervals of $\eta < 1.26\times 10^{15}$ GeV, $-14.77 <\log_{10} A < -14.24$, and $3.60 < \gamma <4.70$, respectively. Again, the results are similar to those of the GCS+SMBHB scenario in the simplified majoron model. Thus, we do not show any separate plots for $10^{-23} \, \text{eV} < m_{\chi} < 10^{-15} \, \text{eV}$ regime in the modified majoron model.

Before concluding this section, one final observation is worth noting. Let us ignore parts of the gravity-motivated arguments of Section~\ref{sec:themodifiedmajoronmodel} and assume that the $U(1)_{B-L}$ local symmetry (and hence the complex scalar $\phi^{\prime}$) is not needed to prevent $d \leq 4$ effective operators induced by gravity. Then, the reduced modified majoron model will only contain an approximate $U(1)$ global symmetry that is also broken spontaneously when the complex scalar $\phi$ attains a VEV $\eta$. In this model, the fit to the latest NANOGrav data will be identical to our results from the complete modified majoron model discussed above for $10^{-23} \, \text{eV} < m_{\chi} < 10^{-15} \, \text{eV}$. However, in this reduced modified majoron model, the fit is valid for all $m_{\chi}$ below $10^{-15}$ eV since, in this scenario, there will be no Type-$\phi^{\prime}$ LCS. The GW spectrum is always given by the GCS-like spectrum of Eq.~(\ref{gw_phi}).

\section{Constraints}
 \label{sec:cons}
In this section, we briefly outline various other cosmological and astrophysical constraints on different $U(1)$ breaking scales.

\subsection{Relativistic degrees of freedom ($\Delta N_{\text{eff}}$)}
\label{sec:DNeff}

 Any additional source of radiation in the early Universe is tightly constrained by the precise measurement of the effective number of relativistic degrees of freedom, $N_{\text{eff}} = 2.99 \pm 0.17$, obtained from the CMB observations by the Planck collaboration~\cite{planck2018}. In the SM, the predicted value of $N_{\text{eff}}$ is 3.046~\cite{Mangano:2005cc}. Hence, from the CMB, we obtain the $2 \sigma$ upper bound on additional relativistic degrees of freedom to be $\Delta N_{\text{eff}} = 0.284$. We ignore the limits on $\Delta N_{\text{eff}}$ from BBN as they are weaker than the CMB. 

\paragraph{The simplest majoron model:} In the simplified majoron model, both GW and majoron, which is a massless Goldstone boson, contribute to the radiation energy density of the Universe. The total relic densities of GWs and Goldstone bosons during the CMB from GCS can be obtained by integrating over the frequency from $f_{in} = 10^{-18}$ Hz to infinity. They can be expressed as $\Delta \Omega_x = \int_{f_{in}}^{\infty} d(\ln f) \Omega_{x}(f)$. In this expression, $x$ stands for the massless majoron and GW radiation from GCS (or from LCS as well in the modified majoron model). The constraint on the total relic density of the radiation species $x$ from the measured $\Delta N_{\text{eff}}$ during the CMB reads as~\cite{Henrot-Versille:2014jua}
\begin{align}
   \Delta \Omega_x h_0^2 \bigg |_{\text{CMB}} \, \leq \, 5.6 \times 10^{-6} \, \Delta N_{\text{eff}} \bigg |_{\text{CMB}} \,  = \, 1.59 \times 10^{-6}\,.
   \label{eq:DNeff_CMB}
\end{align}
Using the Goldstone boson emission spectra from GCSs, as presented in Ref.~\cite{Chang:2021afa}, a $2\sigma$ upper bound on the $U(1)_L$ global symmetry breaking scale is obtained from the CMB data: $\eta < 5.08 \times 10^{15}$ GeV. Here, Ref.~\cite{Chang:2021afa} predicts a $\eta$ dependent spectrum for Goldstone boson emissions from GCSs. However, there are also studies~\cite{Hindmarsh:2019csc, Hindmarsh:2021vih, Buschmann:2019icd, Buschmann:2021sdq}, which suggest an $\eta$ independent Goldstone emission spectra. The same CMB limit on the relic density of GW from GCSs provides a much weaker bound of $\eta < 1.06 \times 10^{16}$ GeV. 

Now, let us present a comparative picture of $\Delta N_{\text{eff}}$ bounds from the CMB and the fit to the NANOGrav data in the simplified majoron model (cf. Table~\ref{tab:gcs}). If GCS is considered as the only source of stochastic GW background, we obtain a $2 \sigma$ range of $2.69\times 10^{15} \, \text{GeV} < \eta < 3.63 \times 10^{15} \, \text{GeV}$. When GCS is only a secondary GW source to a population of SMBHB mergers, we obtain $\eta < 1.26 \times 10^{15}$ GeV, again at $2 \sigma$. So, in explaining the NANOgrav data in the GCS-only scenario, the range of the symmetry-breaking scale $\eta$ just escapes the bound from $\Delta N_{\text{eff}}$ due to majoron emissions. In contrast, the GCS+SMBHB scenario is safe from the present $\Delta N_{\text{eff}}$ bound. However, due to the low BF offered by the above fits, if one assumes these fits as upper limits only, then currently, the NANOGrav 15-year data limits are slightly stronger than those from $\Delta N_{\text{eff}}$. 

The future CMB-S4 experiment is expected measure $N_{\text{eff}}$ with a precision of $\pm 0.07$~\cite{abitbol2019simons}. 
Assuming the central value to be the SM $N_{\text{eff}}$, this will result in a $2 \sigma$ constraint of $\Delta \Omega_x h_0^2 \, \leq 7.8 \times 10^{-7}$ on additional sources of radiation in a BSM model, and will yield $\eta < 3.56 \times 10^{15}$ GeV. Hence, CMB-S4 observations will be in tension with a part of the parameter space in the GCS-only fit to the NANOGrav 15-year data. 

\paragraph{The modified majoron model:} Now, in the modified majoron model, majorons are not massless, and they are pNGB. For light pNGB emissions from Type-$\phi$ GCS-like strings, we do not use the spectrum of Ref.~\cite{Chang:2021afa} as it was derived for massless Goldstone bosons. Instead, we follow the results of Ref.~\cite{Gorghetto:2021fsn}, which performed a dedicated simulation for light pNGB emissions from GCS-like strings, and find
\begin{align}
  \Delta N_{\text{eff}} = 0.6 \bigg( \frac{\eta}{10^{15} \, \text{GeV}}\bigg)^2 \bigg( \frac{\log(m_{\rho}/H_{\text{CMB}})}{90} \bigg)^3 \,. 
\end{align}
Assuming that the mass of the heavy $CP$-even component of $\phi$ satisfies $m_{\rho} \approx \eta$, we obtain a significantly stronger $2\sigma$ bound on the accidental global $U(1)$ symmetry breaking scale, with $\eta < 4.45 \times 10^{14}$ GeV, based on the present CMB data. Instead, an assumption of $m_{\rho} \sim 10^{-3} \eta$ relaxes the bound by only $10 \%$. CMB-S4 will tighten the bound to $\eta < 3.14 \times 10^{14}$ GeV.

We should remind the reader that the GW energy density in this model is dominated by the GW emission from LCSs as this model contains both Type-$\phi^{\prime}$ LCSs and Type-$\phi$ GCS-like strings, and the VEVs satisfy $\eta < \eta^{\prime}$ (for details, please see Section~\ref{gcs+lcs}). Hence, using Eq.~(\ref{eq:DNeff_CMB}) one finds that the Planck $\Delta N_{\text{eff}}$ data places a $2 \sigma$ upper limit of $\eta^{\prime} < 2.07 \times 10^{15}$ GeV on the local $U(1)_{B-L}$ symmetry breaking scale. CMB-S4 will improve this limit to $\eta^{\prime} < 1.06 \times 10^{15}$ GeV.

When Type-$\phi$ and Type-$\phi^{\prime}$ CSs are the only source of stochastic GW background, the fit to the latest NANOGrav data returns $2.95\times 10^{13} \, \text{GeV} < \eta^{\prime} < 6.03 \times 10^{13} \, \text{GeV}$, and $\eta < 1.99 \times 10^{13}$ GeV at $2 \sigma$ for $m_{\chi} < 10^{-23}$ eV. Instead, when CS+SMBHB scenario is considered in this model, the fit yields $1.26 \times 10^{12} \, \text{GeV} < \eta^{\prime} < 5.50 \times 10^{13} \, \text{GeV}$ and  $\eta < 6.31 \times 10^{12}$ GeV for the same range of majoron masses. Hence, clearly, when the global $U(1)$ symmetry is protected by the local $U(1)_{B-L}$ symmetry, the parameter space that fits the NANOGrav 15-year data for $m_{\chi} < 10^{-23}$ eV, evades the present and future $\Delta N_{\text{eff}}$ bounds from the CMB. Again, if one considers the above fits as $2 \sigma$ upper limits only, NANOGrav furnish stronger bounds than $\Delta N_{\text{eff}}$.

Let us now emphasize the importance of protection of the global $U(1)$ symmetry by the local $U(1)_{B-L}$ symmetry in this paragraph. If we disregard the argument that the accidental global symmetry is not completely broken at the Planck scale by gravity, then the local symmetry is not needed, and there will be no Type-$\phi^{\prime}$ LCS. Consequently, the stochastic GW background is provided by Type-$\phi$ GCS-like strings only. In this model, the fit to the NANOGrav data yields $2.75\times 10^{15} \, \text{GeV} < \eta < 3.63 \times 10^{15} \, \text{GeV}$ ($\eta < 1.26\times 10^{15}$ GeV) at $2 \sigma$ for CS-only (CS+SMBHB) scenario. Thus, the explanation of the NANOGrav data in the CS-only scenario in this model is quite evidently ruled out by $\Delta N_{\text{eff}}$ from the CMB. The same conclusion is true for the full $U(1)_{B-L} \times U(1)$ model for majoron mass range $10^{-23} \, \text{eV} < m_{\chi} < 10^{-15} \, \text{eV}$.

\subsection{CMB anisotropies and iso-curvature perturbations}
\label{sec:CMB-ani+Isocurv}
\paragraph{CMB anisotropy:} It is well known that if CSs exist during the epoch of decoupling, their gravitational effects can generate extra anisotropies in the CMB~\cite{Zeldovich:1980gh,Vilenkin:1981iu}. If photons pass through a long CS, then it appears to an observer that a slight Doppler shift has been induced between the photons. As a result, CSs cause abrupt temperature variations in the CMB photons near them, with a characteristic amplitude $\delta T/T$ proportional to $G\mu$~\cite{Kaiser:1984iv}. Because strings behave as random sources, CSs can account for, at most, a modest fraction of the total CMB anisotropy~\cite{Pogosian:2003mz, Wyman:2005tu, Fraisse:2006xc}.

In a nutshell, CMB anisotropy measurements constrain the symmetry breaking scale to $\eta < 6.3 \times 10^{14}$ GeV for GCS~\cite{Lopez-Eiguren:2017dmc} and $\eta^{\prime} < 2.2 \times 10^{15}$ GeV for LCS~\cite{Lizarraga:2016onn}, but these bounds apply only for $m_{\chi} < 10^{-28}$ eV. Consequently, in this mass range, the CS-only fit to the NANOGrav data is excluded for the GCS-like spectrum of the reduced modified majoron model with a single accidental global $U(1)$ symmetry. Also, the majorons are massless in the simplified majoron model. Hence, the GCS-only fit for the pure GCS spectrum of the simplified majoron model is also in tension with CMB anisotropy observations. 
Furthermore, the parameter spaces for scenarios where SMBHB mergers are included with GCSs as GW sources get constrained by CMB anisotropy in both models. Notably, however, the full $U(1)_{B-L} \times U(1)$ modified majoron model remains unaffected by these constraints, preserving its compatibility with the NANOGrav data.

\paragraph{Iso-curvature perturbations:} Models that contain softly broken accidental global $U(1)$ symmetries, DWs will form. When DWs annihilate, they will develop inhomogeneities in the majoron field~\cite{Hogan:1988mp}. Even if majorons form a fraction of the DM, these inhomogeneities can introduce iso-curvature fluctuations, which may be in tension with cosmological observations.

Ref.~\cite{Gorghetto:2021fsn} has evaluated the bounds from iso-curvature perturbations from CMB and Lyman-$\alpha$ observations for axions, which also obtain a mass due to a broken, accidental $U(1)$ global symmetry. However, there are significant uncertainties in deriving these constraints (please see Ref.~\cite{Gorghetto:2021fsn} for details). Nevertheless, if we take these bounds on their face value, CMB (Lyman-$\alpha$) imposes the strongest bound of $\eta \, \text{or} \, \eta^{\prime} <3 \times 10^{14} \, (2.5 \times 10^{14})$ GeV for $m_{\chi} \sim 10^{-29} \, (10^{-22})$ eV. The limits relax sharply on either side of the majoron masses quoted above.

Again, isocurvature bounds exclude the fit to the NANOGrav 15-year data unless the global $U(1)$ symmetry is protected by a local $U(1)_{B-L}$ symmetry. However, when this protection is present, the fit readily evades these constraints for $m_{\chi} < 10^{-23}$ eV.

As has been described in this section, the BSM parameters in the modified majoron model, including the symmetry breaking scales $\eta$ and $\eta^\prime$ as well as the majoron mass, are susceptible to various constraints. Among those constraints, the ones on $\eta$ from CMB anisotropy for GCS, isocurvature perturbation, $\Delta N_{eff}$ for pNGB emissions, and fuzzy DM bounds from Lyman-$\alpha$ have been summarized in Fig.~\ref{fig:DM_fin} \textit{(top)}. Similarly, the constraints on $\eta^\prime$ coming from CMB anisotropy for LCS, isocurvature perturbation, and $\Delta N_{eff}$ for GW emissions from LCS have been encapsulated in Fig.~\ref{fig:DM_fin} \textit{(bottom)}. Finally, we have shown the $2\sigma$ credible intervals of the best fit parameter regions overlaid on those constraints.

Apart from the constraints mentioned above, there exist constraints from the LIGO non-observation of stochastic GWs which can be translated to the vevs associated with this model as $\eta \lesssim10^{17}$ GeV for GCS-scenario, and $\eta^\prime \lesssim10^{16}$ GeV for type-C modified majoron scenario; both of which remain well above the region of our interest.

\section{majoron Dark Matter}
\label{sec:majoronDM}
A typical model of particle physics with a stable electromagnetically neutral BSM particle can provide an opportunity to explore DM, which is one of the biggest mysteries of particle and astroparticle physics that has eluded the scientific fraternity for quite some time. The modified majoron model described in Section~\ref{sec:themodifiedmajoronmodel} is no exception in this regard. 
In this section, we explore whether the majoron in the modified majoron model can possibly be the DM while taking into account constraints from CSs that are consistent with the latest NANOGrav data as discussed in Section~\ref{sec:results} and various other cosmological limits described in Section~\ref{sec:cons}.
The majoron can be produced in the early Universe using several different mechanisms, subject to certain necessary conditions. The production mechanisms that are going to be taken into account in this work are threefold and can be listed as follows: a thermal production from the cosmic soup \cite{Rothstein:1992rh}, through a coherent oscillation of the majoron around its minimum \cite{Preskill:1982cy,Abbott:1982af,Dine:1982ah} and non-thermal radiation from the CSs \cite{Gorghetto:2018myk,Gorghetto:2020qws,Gorghetto:2021fsn}.

The majoron mass range of interest in this analysis is tiny compared to the typical VEVs (i.e., $\eta$ and $\eta^\prime$) required for the CS discussion in light of the NANOGrav data. Thus, the BSM states that have sizable couplings with the majoron are very heavy and decouple early from the cosmic soup, resulting in the departure of the majoron from the equilibrium. The thermalized lighter states, such as the light neutrinos, can still interact with the majoron with a VEV-suppressed strength but cannot bring it back to the thermal soup within the parameter space depicted in this work. Hence, the spontaneous breaking of the high-scale symmetry assures a thermal production of majoron, which remains a DM relic as long as it satisfies the stability criteria. 
Among the three light neutrino species, majoron decay into the heavier two is kinematically forbidden. Although the couplings between the majoron and light neutrinos are highly suppressed, decay through the lightest neutrino remains possible if kinematically allowed since the lightest neutrino mass can, in principle, be smaller than the majoron mass. The lifetime of the majoron in such a decay channel is given by
\begin{equation}
\tau_\chi \sim 3 \times 10^{89} \left(\frac{\eta}{10^{13} \,\textrm{GeV}}\right)^2 \left(\frac{10^{-16} \,\textrm{eV}}{m_\nu}\right)^2 \left(\frac{10^{-15} \,\textrm{eV}}{m_\chi}\right)  \, \textrm{sec}.
\label{eq:lifetime_DM}
\end{equation}
This lifetime far exceeds the age of the universe, thereby ensuring the majoron's stability on cosmological timescales as a DM candidate.
The DM relic for this thermal component is proportional to the majoron mass
{
\begin{equation}
\Omega_{\chi,th}h^2\approx1.15\times10^{-18}\bigg(\frac{m_\chi}{10^{-15}\textrm{ eV}}\bigg) \,.
\label{eq:therm_DM}
\end{equation}}

The gravity-induced higher-dimensional operator mentioned in Eq.~(\ref{eq.4}) can break the global symmetry explicitly, leading to a coherent oscillation of the majoron field around its minimum. It has been shown in the above-mentioned references \cite{Preskill:1982cy,Abbott:1982af,Dine:1982ah} that the oscillation energy may achieve comparable energy densities to the thermal component, especially in the presence of $\mathcal{O}(10)$ and higher dimensional operators in the Lagrangian. We have adopted an $\mathcal{O}(1)$ angle parameter signifying the initial misalignment of the Goldstone field with the direction of the explicit symmetry breaking for our analysis following Ref.~\cite{Rothstein:1992rh}. Since the wavelength of such oscillations is higher than the Compton wavelength, it can be treated as non-relativistic DM with a relic given by {
\begin{equation}
\Omega_{\chi,osc}h^2\approx7\times10^{-6}g_*^{\frac{3}{4}}\bigg(\frac{m_\chi}{10^{-15}\textrm{ eV}}\bigg)^{\frac{1}{2}}\bigg(\frac{\eta_{eff}}{10^{13}\textrm{ GeV}}\bigg)^2\frac{qq^\prime}{nn^\prime}
\label{eq:osc_DM}
\end{equation}}
where $g_*$ (an $\mathcal{O}(1)$ number) is the number of equilibrated degrees of freedom at $T_*$ such that $m_\chi=3H(T_*)$. The stability of such a DM component is implicit in the operator dimension mentioned above in this paragraph.

The CSs emit majorons continuously throughout the scaling regime. Once majoron becomes nonrelativistic as the Hubble parameter decreases roughly down to $m_\chi$, it can act as a DM candidate subject to stability. A similar analysis has been conducted in Ref.~\cite{Gorghetto:2021fsn} in the context of axions, which can be followed to arrive at the following expression for DM relic
{
\begin{equation}
\Omega_{\chi,str}h^2\approx1.4\times10^{-2}\bigg(\frac{\eta}{10^{13}\textrm{ GeV}}\bigg)^2\bigg(\frac{m_\chi}{10^{-15}\textrm{ eV}}\bigg)^\frac{1}{2} \, ,
\label{eq:str_DM}
\end{equation}}
where the nonlinear transient effect \cite{Gorghetto:2020qws} and the relativistic redshift of majorons after $H \sim m_\chi$ have not been taken into account. The cumulative effect leads to a non-conservation of the number density of the majoron, which is, however, not expected to introduce a change in the order of magnitude of the relic abundance for the chosen parameter space and is hence neglected. {Also, additional majorons are expected to be produced after the DW formation and during the collapse of the string-wall network. However, we do not take this component into account as a reliable calculation and prediction of majoron production during the destruction of the network is not available in the literature.}

\begin{table}[tp!]
\centering

\renewcommand{\arraystretch}{1.6} 

\begin{tabular}{|c|c|c|c|c|c|c|c|c|c|}\hline
\hspace{4.5pt}Benchmarks\hspace{4.5pt} & \hspace{5.5pt}$n$\hspace{5.5pt} & \hspace{5.5pt}$n^\prime$\hspace{5.5pt} & \hspace{5.5pt}$d$\hspace{5.5pt} & $\eta$ (GeV) & $\eta^\prime$ (GeV) & $g_{grav}$ & $m_\chi$ (eV)  &\hspace{4.5pt}$\Omega_{DM}h^2$\hspace{4.5pt} \\
\hline\hline
BP 1 & 1 & 15 & 16 & \hspace{4.5pt}$4.9\times10^{13}$\hspace{4.5pt} & \hspace{4.5pt}$5.0\times10^{13}$\hspace{4.5pt}  & \hspace{4.5pt}$6.3\times10^{-16}$\hspace{4.5pt} & \hspace{4.5pt}$1.2\times10^{-16}$\hspace{4.5pt}  &0.1205 \\
\hline
BP 2 & 1 & 16 & 17 & $2.9\times10^{13}$ & $5.0\times10^{13}$ & $1.2\times10^{-8}$ & $9.3\times10^{-16}$  & 0.1192 \\
\hline
\end{tabular}
\caption{A few representative benchmark points consistent with the constraints mentioned in Sec.~\ref{sec:cons} that can satisfy the measured DM relic abundance.}
\label{tab:BP_DM}
\end{table}

The total DM relic can be arrived at by combining the components from Eqs.~(\ref{eq:therm_DM}), (\ref{eq:osc_DM}) and (\ref{eq:str_DM}) which would depend on various model parameters listed as $d$, $n$, $n^\prime$, $q$, $q^\prime$, $\eta$, $\eta^\prime$, $\eta_{eff}$, $m_\chi$ and $g_{\textrm{grav}}$. Among these parameters, $\eta_{eff}$ and $m_\chi$ can be written in terms of the other parameters following Eqs.~(\ref{eq.majoronMass}) and (\ref{eq:etaeff}). The value of $q$ has been fixed to 2 in Section~\ref{sec:themodifiedmajoronmodel} and the $q^\prime$ can be expressed in terms of $q$, $n$ and $n^\prime$ utilizing conservation of the $B-L$ charge in the higher dimensional operator of the lagrangian mentioned in Eq.~(\ref{eq.4}). The value of $\eta^\prime$ is fixed roughly at $\mathcal{O}(10^{13})$ GeV after taking the data of Table~\ref{tab:cs_mmm} into consideration {when a combination of Type-$\phi^{\prime}$ LCS and Type-$\phi$ GCS-like strings fit the NANOGrav data.} Furthermore, one of the parameters among $n$ and $n^\prime$ can be traded off with the other and $d$ using dimensional analysis of the lagrangian term in Eq.~(\ref{eq.4}). That leaves us with the parameters $d$, $\eta$, $g_\textrm{grav}$ and $n$ (or $n^\prime$) that can be varied independent of each other. In what follows, we shall try to constrain this four-dimensional parameter space using the DM relic abundance using Planck data ($\Omega_{DM}h^2=0.120\pm0.001$\cite{Planck:2018vyg}). 

\begin{figure}[tp!]
\centering
\includegraphics[width=0.499\textwidth]{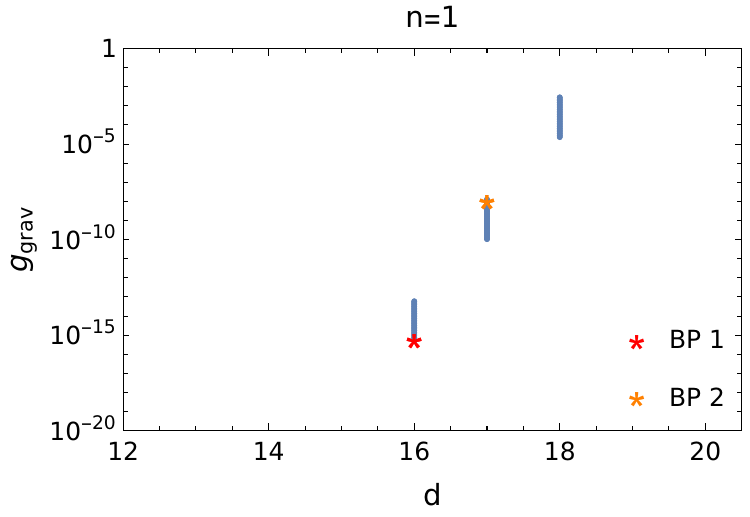}~\includegraphics[width=0.499\textwidth]{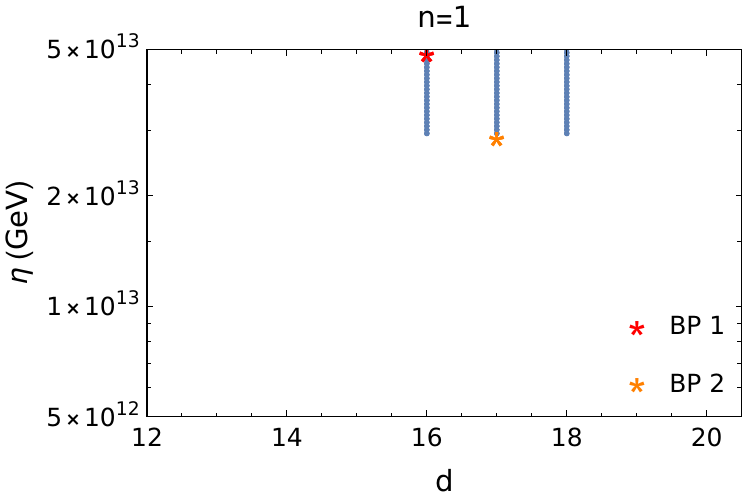}
\caption{Data points that explain the dark matter relic abundance for $\eta^\prime=5\times10^{13}$ GeV and $n=1$.}
\label{fig:DM_scan}
\end{figure}

\begin{figure}[tp!]
\centering
\includegraphics[width=0.5\textwidth]{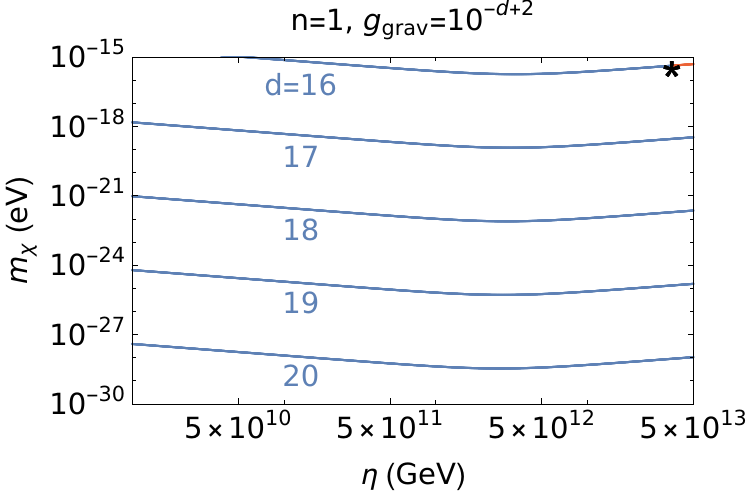}
\caption{Data points that explain dark matter relic abundance for $\eta^\prime=5\times10^{13}$ GeV, $n=1$ and $g_\textrm{grav}=10^{-d+2}$. The blue and red dots represent the underabundant and overabundant regions respectively while the black stars represent the measured DM relic.}
\label{fig:DM_d}
\end{figure}

The analysis is carried out under the assumption, $n=1$ and the corresponding results are displayed. The parameter space is scanned in the following ranges: $d$ in range [5, 20], $\eta$ in range [$10^6$, $5\times10^{13}$] GeV and $g_\textrm{grav}$ in range [$10^{-d}$, 1]~\footnote{
The smallness of $g_\textrm{grav}$ is quite natural for one scalar field $\phi$ breaking a global $U(1)$ symmetry as argued in Section~\ref{sec:themodifiedmajoronmodel}. We do not explicitly show the same for the full $U(1) \times U(1)$ modified majoron model. However, as this smallness arises from the exponential suppression by the combined action ($S$) of the scalar field and gravity, it is natural to assume tiny values of $g_\textrm{grav}$ in the complete modified majoron model as well.}
for explaining the measured DM relic. The parameter space has been restricted to $m_\chi<10^{-15}$ eV in order to maintain the desired GW spectrum {from Type-$\phi$ CSs} in the nano-Hz frequency range. The lowest operator dimension that can satisfy the correct relic within the above-mentioned range is $d=16$ for the $n=1$ case, as shown in Fig.~\ref{fig:DM_scan}. A few representative benchmark points that are consistent with the constraints mentioned in Section~\ref{sec:cons} and can also satisfy the correct DM relic abundance have been mentioned in Table~\ref{tab:BP_DM} and further highlighted in Fig.~\ref{fig:DM_scan}. Next, we fix the $g_\textrm{grav}$ at $10^{-d+2}$ and plot the relic contour in the $\eta$ vs $m_\chi$ plane in Fig.~\ref{fig:DM_d}. In this figure, the points that correspond to correct relic, underabundant relic, and overabundant relic are given, respectively, by black, blue and red colours. The adjacent points represent the same value of $d$.


If a sizable portion of the DM particles stream with sufficiently high velocities, it contradicts the observed structure formation. {For the parameter space under consideration here, Lyman-$\alpha$ observations place limits on how much of the DM can be made up of particles with masses below approximately $10^{-20}$ eV~\cite{Irsic:2017yje, Kobayashi:2017jcf}. This limit is valid for the modified majoron model for $\eta < 7 \times 10^{14}$ GeV~\cite{Gorghetto:2021fsn}, which is weaker than any of the other constraints discussed in Section~\ref{sec:cons}.}

{A careful observation of Table~\ref{tab:BP_DM} and Fig.~\ref{fig:DM_fin} will reveal that when we fix $\eta^{\prime} = 5 \times 10^{13}$ GeV the correct DM relic density is obtained for $\eta \sim \eta^{\prime}$ and $m_{\chi} \sim  10^{-16} - 10^{-15}$ eV. Below these values of $\eta$ and $m_{\chi}$, majoron DM is under-abundant. Our choice of $\eta^{\prime} = 5 \times 10^{13}$ emanated from the fit to the NANOGrav data when both Type-$\phi^{\prime}$ and Type-$\phi$ strings contribute to the GW spectrum in this model. However, to make sure that the Type-$\phi^{\prime}$ LCSs survive long enough so that the resulting GW spectrum reaches below nano-HZ, one requires $m_{\chi}<10^{-23}$ eV in the modified majoron model. Hence, clearly, for the scenario when the latest NANOGrav data is fit by both Type-$\phi^{\prime}$ and Type-$\phi$ strings, which evades all the bounds from Section~\ref{sec:cons}, majoron DM will be significantly under-abundant. They can only form a small fraction of the total DM.}

{In contrast, if one considers $10^{-23} \, \text{eV} <m_{\chi} < 10^{-15} \, \text{eV}$, then the NANOGrav data still can be fit by GW emitted by only Type-$\phi$ strings (see Sections~\ref{gcs+lcs} and \ref{sec:results} for details). However, in such a scenario we require $\eta \sim \mathcal{O}(10^{15})$ GeV. We have not incorporated $\eta$ values up to $\mathcal{O}(10^{15})$ GeV in our scan in this section since this parameter space is in tension with bounds from $\Delta N_{\text{eff}}$, CMB anisotropies, and isocurvature perturbations. However, we should mention here that among the three contributions to the DM relic mentioned in Eqs.~(\ref{eq:therm_DM}), (\ref{eq:osc_DM}), and (\ref{eq:str_DM}), $\Omega_{\chi,str}h^2$ happens to be the dominant one for $m_{\chi} < 10^{-15}$ eV, and one can neglect the other contributions. Equation~(\ref{eq:str_DM}) implies a simple relation, $\Omega_{\chi,\text{str}} h^2 \propto \sqrt{m_{\chi}}  \eta^2$. Using this expression, we plot the contour corresponding to the observed DM relic abundance, $\Omega_{\chi}/\Omega_{\text{DM}} = 1$ (shown in pink), extending up to $\eta = 10^{16}$ GeV in the $\eta$–$m_{\chi}$ plane of Fig.~\ref{fig:DM_fin}. 
}

\section{Summary and conclusion}
\label{sec:conc}
\begin{figure}[tp!]
\centering
\includegraphics[scale=0.9]{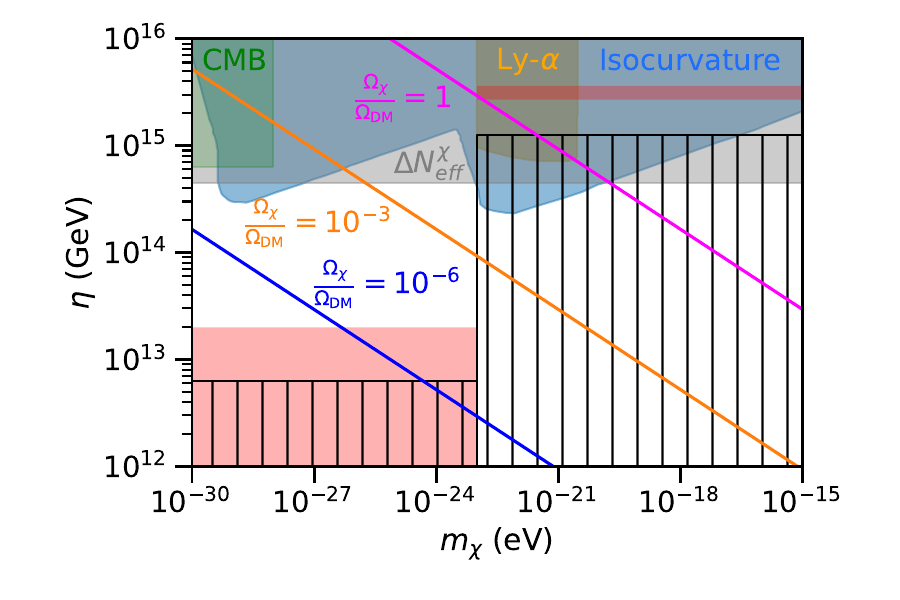}\\
\includegraphics[scale=0.9]{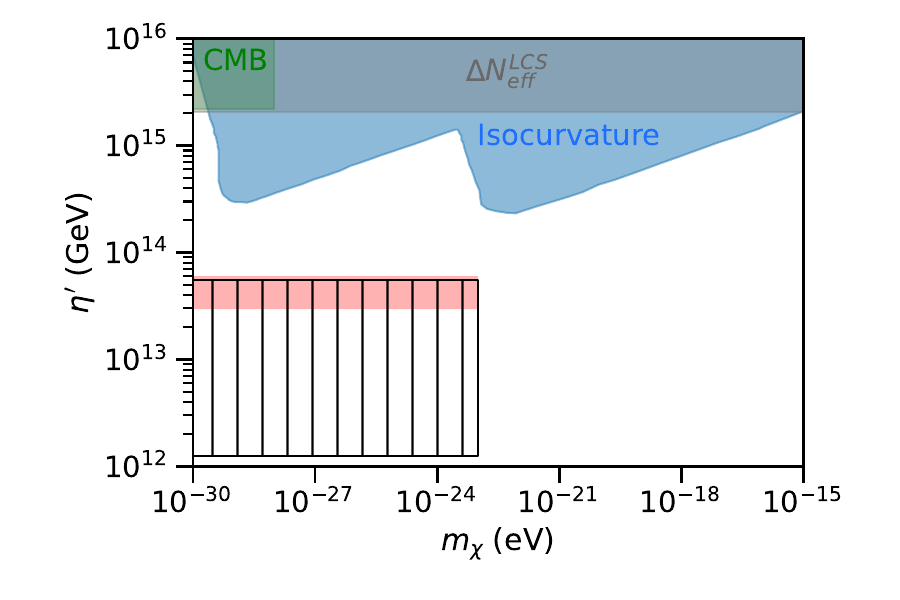}
\caption{The majoron relic contours as a fraction of the observed DM relic for various values \{$10^{-6}$, $10^{-3}$, 1\} are shown on the majoron mass vs $\eta$ parameter space. The light green shaded regions are disallowed from CMB anisotropy measurements for GCS \textit{(top)} and LCS \textit{(bottom)}. The gray shaded region is disallowed by the Planck constraint on $\eta$ and $\eta^\prime$ respectively from pNGB emission \textit{(top)} and GW emission \textit{(bottom)}. Further, the excluded region of the parameter space from isocurvature perturbation is shaded in blue (both \textit{top} and \textit{bottom}) and the orange shaded portion is disallowed by fuzzy DM constraints from Lyman-$\alpha$ \textit{(top)}. For more details on the constraints see the text in Section~\ref{sec:cons}. The $2\sigma$ credible intervals (see Table \ref{tab:gcs} and \ref{tab:cs_mmm}) of parameter space that fit the NANOGrav data for the Type-C case without (with) SMBHB are displayed in light red shaded region (black vertical stripes) in both panels.}
\label{fig:DM_fin}
\end{figure}


In this work, we investigate majoron cosmology in light of the recent evidence for a stochastic GW background reported by the NANOGrav collaboration. The simplest majoron model extends the SM with a global $U(1)_L$ symmetry, three RHNs, and a complex scalar singlet $\phi$, whose VEV spontaneously breaks the symmetry and generates RHN Majorana masses, with the massless phase of $\phi$ identified as the majoron. However, gravitational effects are known to explicitly break global symmetries at the Planck scale due to the formation of wormholes, and integrating wormholes out leads to an effective low-energy theory given by Eq.~(\ref{eq:L_wh_eff})~\cite{Abbott:1989jw}. Thus, if the global $U(1)$ symmetry emerges accidentally at the see-saw scale and is softly broken by gravity, it allows the majoron to survive as a pNGB and potential dark matter candidate - unlike in local $U(1)_L$ or $U(1)_{B-L}$ see-saw scenarios.

Again, referring back to the wormhole-induced effective Lagrangian of Eq.~(\ref{eq:L_wh_eff}), it is not rigorously justified that a single scalar is sufficient to suppress all effective operators with dimension $d \leq 4$, and thereby prevent explicit hard breaking of the global symmetry at the Planck scale. Besides, the fit to the NANOGrav data when one global $U(1)$ symmetry is softly broken by gravity is in tension with bounds from  $\Delta N_{\text{eff}}$, CMB anisotropies, and isocurvature perturbations (see Section~\ref{sec:cons}). 

Hence, to ensure gravity-induced $d \leq 4$ operators are forbidden, a more conservative model is adopted by gauging $U(1)_{B-L}$ and introducing an additional scalar $\phi^{\prime}$~\cite{Rothstein:1992rh} on top of the model discussed above. Hence, the model extends the SM gauge group by a local $U(1)_{B-L}$ symmetry and includes an approximate global $U(1)$ symmetry. Both $\phi$ and $\phi^{\prime}$ are charged under $U(1)_{B-L}$, while only $\phi$ carries the global charge. Assigning $B-L$ charge of $q=2$ to $\phi$ enables Majorana mass generation for RHN. In contrast, the $B-L$ charge $q^{\prime}$ is left general to enable model building for various $d$-dimensional soft global symmetry-breaking operators by the Lagrangian term of Eq.~(\ref{eq.4}). A few phenomenologically motivated benchmark models are shown in Table~\ref{tab:BP_DM}. We refer to this $U(1)_{B-L} \times U(1)$ model as the modified majoron model. 

While our primary focus in this paper is the cosmology of the modified majoron model, we also examine the corresponding features of the simplest majoron model. The key observations of our study are summarized below.

\vspace{0.2 in}

\paragraph{The simplest majoron model:}
\begin{itemize}
    \item The spontaneous breaking of the global $U(1)_L$ symmetry leads to GCSs that produce a stochastic GW background.
     
     \item With GCS-only GW source, the NANOGrav 15-year data yields a $2\sigma$ credible interval of $2.69 \times 10^{15} < \eta < 3.63 \times 10^{15}$ GeV (see Table~\ref{tab:gcs} and Fig.~\ref{fig:GCS_posterior}). However, this fit is disfavored relative to the SMBHB-only scenario, with a BF of $7.66 \times 10^{-5}$. 
     When GCS and SMBHB sources are combined, the GW fit improves with BF $\sim 0.65$. The inferred $2\sigma$ upper bound on the symmetry-breaking scale is $\eta < 1.26 \times 10^{15}$ GeV (see Table~\ref{tab:gcs} and Fig.~\ref{fig:GCS+SMBHB_posterior}).
     The fits constrain the neutrino-sector coupling $\delta_{33}$ (cf. Eq.~(\ref{eq.3})) to the range $12.57$--$16.97$ in the GCS-only scenario and to $\delta_{33} < 5.89$ in the GCS+SMBHB case.
     
     \item $\Delta N_{\text{eff}}$ bounds on the relic densities of Goldstone bosons and GWs from GCSs lead to $2\sigma$ upper limits of $\eta < 5.08 \times 10^{15}$ GeV (from majoron emission) and a weaker $\eta < 1.06 \times 10^{16}$ GeV (from GW emission). CMB anisotropy constraints impose upper bounds on the symmetry breaking scale $\eta < 6.3 \times 10^{14}$ GeV for GCSs (see Section~\ref{sec:cons}). When GCS is considered the only source of GW, the explanation for the NANOGrav 15-year data is in tension with the CMB anisotropy measurement.
\end{itemize}

\vspace{0.2 in}

\paragraph{The modified majoron model:}

\begin{itemize}

    \item The modified majoron model features sequential symmetry breaking: $U(1)_{B-L}$ (local) breaks at $T \sim \eta^{\prime}$ forming Type-$\phi^{\prime}$ LCSs, followed by global $U(1)$ breaking at $T \sim \eta$ generating Type-$\phi$ GCS-like strings. At $T \sim T_*$ (set by $m_\chi = 3H(T_*)$), DWs appear due to the explicit symmetry breaking by the operator in Eq.~(\ref{eq.4}). 

    \item The number of DWs attached to each string is determined by the winding numbers and the structure of the symmetry-breaking operator of Eq.~(\ref{eq.4}). When one of the conditions of Eq.~(\ref{eq:oneDW}) is satisfied, only one DW is attached to each string, ensuring the collapse of the string-wall system and avoiding the `domain wall problem'. In this paper, we only consider the $n=1$ condition.

    \item In this model, the total GW signal (Eq.~(\ref{eq.total}) and Table~\ref{t.model}) arises from independent contributions of Type-$\phi^{\prime}$ LCSs and Type-$\phi$ GCS-like strings, distinguishing it from simpler majoron scenarios. The IR cut-off in the Type-$\phi$ spectrum scales as $\sqrt{m_{\chi}}$, making it valid down to nano-Hz frequencies only if $m_\chi \lesssim 10^{-15}$ eV. The Type-$\phi^{\prime}$ LCS spectrum remains within the PTA band for $m_\chi \lesssim 10^{-23}$ eV (assuming $\eta^{\prime} \sim 10^{14}$ GeV). Thus, for $m_\chi < 10^{-23}$ eV, both components contribute to the NANOGrav-observable GW background, whereas for $10^{-23} \, \text{eV} < m_\chi < 10^{-15} \, \text{eV}$, only the Type-$\phi$ spectrum extends below nano-Hz.
    
    \item For $m_\chi \lesssim 10^{-23}$ eV and CS-only GW spectrum, the NANOGrav 15-year data is best fit with \\${2.95 \times 10^{13} \, \text{GeV} < \eta^{\prime} < 6.03 \times 10^{13}}$ GeV and ${\eta < 1.99 \times 10^{13}}$ GeV at $2\sigma$ (Fig.~\ref{fig:fig3} and Table~\ref{tab:cs_mmm}). The CS+SMBHB combined fit yields ${1.26 \times 10^{12} \, \text{GeV} < \eta^{\prime} < 5.50 \times 10^{13}}$ GeV and ${\eta < 6.31 \times 10^{12}}$ GeV (Fig.~\ref{fig:figx} and Table~\ref{tab:cs_mmm}). The effective neutrino-sector coupling is constrained to $\delta_{33} < 0.09 \, (0.03)$ in the CS-only (CS+SMBHB) scenario. The CS+SMBHB posterior (Fig.~\ref{fig:figx}) shows two distinct regions: one where SMBHB dominates and another where Type-$\phi^{\prime}$ strings are the primary GW source. However, low BF values in both scenarios indicate that the SMBHB merger scenario still provides a better fit to the NANOGrav 15-year data.

    \item For $10^{-23} \, \text{eV} < m_{\chi} < 10^{-15} \, \text{eV}$, only Type-$\phi$ GCS-like strings contribute, with the fit giving ${2.75 \times 10^{15} < \eta < 3.63 \times 10^{15}}$ GeV (CS-only), and ${\eta < 1.26 \times 10^{15}}$ GeV (CS+SMBHB), which are almost identical with the simplified majoron model.
    
    \item For the modified majoron model with pNGB majorons, CMB $\Delta N_{\text{eff}}$ data imposes strong $2\sigma$ upper limits on the VEVs: $\eta < 4.45 \times 10^{14}$ GeV for the global symmetry and $\eta' < 2.07 \times 10^{15}$ GeV for the local $U(1)_{B-L}$ symmetry. CMB anisotropy measurements constrain the symmetry breaking scale to $\eta < 6.3 \times 10^{14}$ GeV for Type-$\phi$ and $\eta^{\prime} < 2.2 \times 10^{15}$ GeV for Type-$\phi^{\prime}$ strings, but these bounds apply only for $m_{\chi} < 10^{-28}$ eV. Iso-curvature perturbations from CMB (Lyman-$\alpha$) observations impose $\eta \, \text{or} \, \eta^{\prime} <3 \times 10^{14} \, (2.5 \times 10^{14})$ GeV for $m_{\chi} \sim 10^{-29} \, (10^{-22})$ eV. These constraints are shown in Fig.~\ref{fig:DM_fin} along with the parameter spaces that fit the NANOGrav 15-year data.

    \item The presence of a local $U(1)_{B-L}$ symmetry protects the global $U(1)$ symmetry for $m_\chi \lesssim 10^{-23}$ eV, allowing the NANOGrav-preferred parameter space to evade stringent CMB and other early-Universe constraints. However, for $10^{-23} \, \text{eV} < m_{\chi} < 10^{-15} \, \text{eV}$, the fit is ruled out by all the above bounds (Fig.~\ref{fig:DM_fin}).

    \item The majoron in the modified majoron model can serve as a viable DM candidate, with its relic abundance arising from three production mechanisms: thermal freeze-out, coherent oscillations (misalignment), and non-thermal radiation from CSs (Eqs.~(\ref{eq:therm_DM})–(\ref{eq:str_DM})). In the vicinity of the parameter space of our concern, correct relic abundance is achieved for operator dimension $d = 16$, majoron masses $m_\chi \sim 10^{-16}$–$10^{-15}$ eV, and $\eta \sim 3–5 \times 10^{13}$ GeV, for both $n=1$ and $n^{\prime}=1$ (Figs.~\ref{fig:DM_scan} and \ref{fig:DM_d}). The scenario where both Type-$\phi$ and Type-$\phi'$ CSs fit the NANOGrav data (requiring $m_\chi < 10^{-23}$ eV) leads to majoron being a subdominant DM component. However, in the range $10^{-23} \,\text{eV} < m_\chi < 10^{-15}$ eV, the observed relic can be satisfied and the NANOGrav data can be matched using only Type-$\phi$ strings—albeit at the cost of tension with CMB and isocurvature bounds (Fig.~\ref{fig:DM_fin}).    
\end{itemize}

In conclusion, the $U(1)_{B-L} \times U(1)$ modified majoron model, with GW contributions from Type-$\phi$ GCS-like and Type-$\phi^{\prime}$ LCSs, can explain the NANOGrav data for $m{\chi} < 10^{-23}$ eV, generate light neutrino masses, and evade cosmological bounds - although the majoron remains a subdominant component of DM. If the NANOGrav fits are interpreted as constraints due to low BFs, they provide limits that are significantly stronger than those from CMB and other cosmological observations.


\acknowledgments
The authors are grateful to express their gratitude to the Department of Atomic Energy
(DAE), India, for the support of Harish-Chandra Research Institute. The authors would also like to thank Ashoke Sen, Anirban Basu, Santosh Kumar Rai, Mairi Sakellariadou, Tanmay Vachaspati, K.S. Babu and Utsav Atta for useful comments and discussions.

\appendix

\section{GW energy spectrum from SMBHB mergers}
\label{sec:smbhb}

For the purpose of fitting a PTA dataset using a power-law spectrum, the GW relic energy density from SMBHB mergers is given by~\cite{Mitridate:2023oar}
\begin{align}
h^2 \Omega_{\text{GW}}(f)  = \frac{2 \pi^2}{3} \frac{1}{H_0^2} A^2 \bigg( \frac{f}{\text{year}^{-1}} \bigg)^{5- \gamma} \, \text{year}^{-2}
 \,,
\label{equation:relic}
\end{align}
where, $\text{H}_0$ is the Hubble expansion rate today, and $h$ is the reduced Hubble constant. $A$ and $\gamma$ denote the amplitude and spectral index, respectively.

By setting the switch \texttt{bhb\_th\_prior=True} in \texttt{PTArcade}, we use a $2D$ gaussian prior for $(\log_{10}A, \gamma)$ in this paper. The mean and covariance matrices for this bivariate distribution are given by
\begin{align}
\mu = \begin{pmatrix}
-15.6 \\
4.7 
\end{pmatrix}
 \,, \, \sigma = \begin{pmatrix}
2.8 & -0.026 \\
-0.026 & 1.2 
\end{pmatrix} 
\,.
\label{eq:SMBHB_prior}
\end{align}

\bibliographystyle{apsrev4-2}
\bibliography{Ref_Paper}
\end{document}